\documentclass[%
 superscriptaddress,
 preprint,
 amsmath,amssymb,
 aps,
 showkeys,
 prx
]{revtex4-1}

\usepackage{graphicx}
\graphicspath{ {../Graphics/} }
\usepackage{dcolumn}
\usepackage{bm}
\usepackage{soul}

\begin{document}

\preprint{APS/123-QED}

\title{Characterising optical fibre transmission matrices using metasurface reflector stacks for lensless imaging without distal access}

\author{George S.D. Gordon}
	\affiliation{Department of Electrical and Electronic Engineering, University of Nottingham}
	\email{George.Gordon@nottingham.ac.uk}
\author{Milana Gataric}
	\affiliation{Department of Pure Mathematics and Mathematical Statistics, University of Cambridge}
\author{Alberto Gil C. P. Ramos}
	\affiliation{Nokia Bell Labs, Cambridge, UK}
\author{Ralf Mouthaan}
	\affiliation{Electrical Engineering Division, Dept. of Engineering, University of Cambridge}
	\affiliation{Department of Physics, University of Cambridge}
\author{Calum Williams}
	\affiliation{Electrical Engineering Division, Dept. of Engineering, University of Cambridge}
\author{Jonghee Yoon}
	\affiliation{Department of Physics, University of Cambridge}
\author{Timothy D. Wilkinson}
	\affiliation{Electrical Engineering Division, Dept. of Engineering, University of Cambridge}
\author{Sarah E. Bohndiek}
	\affiliation{Department of Physics, University of Cambridge}
	\affiliation{CRUK Cambridge Institute, University of Cambridge}

\date{\today}

\begin{abstract}
The ability to form images through hair-thin optical fibres promises to open up new applications from biomedical imaging to industrial inspection.  Unfortunately, their deployment has been limited because small changes in mechanical deformation (e.g. bending) and temperature can completely scramble optical information, which distorts the resulting images. Since such changes are dynamic, correcting them requires measurement of the fibre transmission matrix \emph{in situ} immediately before imaging.  Transmission matrix calibration typically requires access to both the proximal and distal facets of the fibre simultaneously, which is not feasible during most realistic usage scenarios without compromising the thin form factor with bulky distal optics.  Here, we introduce a new approach to determine the transmission matrix of multi-mode or multi-core optical fibre in a reflection-mode configuration without requiring access to the distal facet.  A thin stack of structured metasurface reflectors is used at the distal facet of the fibre to introduce wavelength-dependent, spatially heterogeneous reflectance profiles. We derive a first-order fibre model that compensates these wavelength-dependent changes in the fibre transmission matrix and show that, consequently, the reflected data at 3 wavelengths can be used to unambiguously reconstruct the full transmission matrix by an iterative optimisation algorithm.  We then present a method for sample illumination and imaging following reconstruction of the transmission matrix. Unlike previous approaches, our method does not require the fibre matrix to be unitary making it applicable to physically realistic fibre systems that have non-negligible power loss.  We demonstrate the transmission matrix reconstruction and imaging method first using simulated non-unitary fibres and noisy reflection matrices, then using much larger experimentally-measured transmission matrices of a densely-packed multi-core fibre. Finally, we demonstrate the method on an experimentally-measured multi-wavelength set of transmission matrices recorded from a step-index multimode fibre. Our findings pave the way for online transmission matrix calibration \emph{in situ} in hair-thin optical fibres.
\end{abstract}

\maketitle


\section{Introduction}
Lensless imaging through hair-thin optical fibres is a technique that promises to open up new areas in biomedical imaging such as: \emph{in vivo} bright-field, dark-field, and fluorescence microscopy \cite{Cizmar2012}; quantitative phase and polarimetric imaging with applications in early detection of cancer \cite{Gataric2018, Gordon2018}; and endoscopic confocal microscopy for high-resolution imaging \cite{Loterie2015a}. For all these applications, accurate characterisation of the deterministic propagation of light through the fibre, discretised as the transmission matrix (TM) \cite{Rotter2017}, is essential for imaging, whether used directly \cite{Gordon2018}, indirectly via phase conjugation \cite{Dunning1982} or via optimisation \cite{Leonardo2011}.  

Unfortunately, the TM is highly sensitive to small changes in mechanical deformation and temperature that are unavoidable in living subjects, so the TM should be calibrated immediately before imaging. The most direct way of measuring a TM is to send pre-defined optical fields into the distal facet of the fibre (furthest from the operator) and measure the resulting fields at the proximal facet (nearest the operator). This widely-used approach has a critical limitation: the distal facet of the fibre is deep within the subject during imaging so the pre-defined optical fields cannot be reliably generated without additional bulky distal optics that would compromise the ultra-thin form factor.

Several methods have been proposed to overcome this limitation and open new avenues for minimally invasive optical imaging in inaccessible areas of the body, e.g. deep in the brain where the fibre may need to curve for safe access \cite{Vasquez-Lopez2018}. Placing a holographic plate on the distal facet of the fibre, which is illuminated via a physically separate single-mode fibre, can be used to create a `virtual beacon' \cite{Farahi2013}. A phase-conjugated version of the resultant proximal field is used to recreate a single focussed spot at the distal facet.  This spot is typically fixed in position but with a multi-core fibre (MCF), the `memory effect' could be exploited to enable scanning over a small distances \cite{Stasio2015}. This method is only able to partially recover the TM and so while confocal imaging is possible, wide-field imaging (e.g. quantitative phase) is not achievable.  Further, the addition of a single-mode fibre to illuminate the beacon adds bulk to the system.

Highly accurate modelling of TM perturbations has also been shown to be feasible, but requires precision manufactured fibres such as telecommunications OM4 grade multi-mode fibre (MMF), is limited to short distances ($<$10cm) \cite{Ploschner2015} and needs very precise bending characterisation, for example through the addition of a Bragg-grating shape sensing fibre \cite{Spillman2014,Udd2011}. Another recently demonstrated method uses reflectors at the distal end to introduce time delays to resolve the fibre TM \citep{Chen2018}. However, this adds undesirable complexity to the system and also requires a unique reflector for each characterised propagation mode, which does not scale well for high-resolution imaging applications.

Gu et al. propose a more generic approach to \emph{in situ} TM characterisation: to infer the forward TM based on the reflection matrix (RM) i.e. light that has made a round-trip into the proximal facet, off a distal reflector and back out the proximal facet \cite{Gu2015}. Applied naively, this approach suffers two major limitations. The first arises due to the transpose symmetry of fibre TMs, which means that measured RMs exhibit a quadratic relationship with their respective TMs, resulting in ambiguities during recovery \cite{Lee2018a, Gu2015}.  In some cases this ambiguity reduces to an $N$-dimensional vector of sign errors (i.e. a vector $\in \{-1,1\}^N$) for an $N$-pixel image, for example, when using MCF with low core-to-core coupling \cite{Warren2016} or using an MMF that is near perfectly unitary (via the Autonne-Takagi matrix factorisation) \cite{Gu2015}.  In such cases, the ambiguity may be resolved by \emph{in situ} optimisation assuming prior knowledge of the sample \cite{Warren2016} or by treating the ambiguities as invariants of the fibre that can be measured in advance \cite{Gu2015}. In practical cases, however, TMs are not unitary even in precision-manufactured MMF \cite{Gu2017, Carpenter2014} because high-order modes, with more power concentrated in the fibre cladding, tend to exhibit power leakage at bends \cite{Gordon2014}. 

The second limitation arises because light must be able to pass through the distal facet of the fibre to enable imaging.  Gu et al. address this by proposing to include a shutter at the distal facet, but again, this added bulk would compromise the ultra-thin form factor \cite{Gu2015}.

Seeking to overcome the need for precision-manufactured MMFs, while maintaining the hair-thin property of the fibres, we demonstrate here a new method that enables measurement of non-unitary fibre TMs of arbitrary size without access to the distal facet and without adding distal bulk, making it suitable for ultra-thin flexible imaging devices. The key innovation is a distal reflector comprising a multi-layer stack of spatially heterogeneous partially reflecting wire-grid polarisers (termed `metasurfaces') and long-pass optical filters.  The filter stack creates a spatially and spectrally varying reflector, whose properties can be characterised prior to use and will remain fixed throughout any imaging experiment.  By combining multiple RMs recorded at different wavelengths it is possible to estimate the instantaneous TM and then perform imaging at another wavelength.

We first introduce a new method that analytically extracts the TM based on 3 measured RMs at different wavelengths and prior knowledge of the reflector properties (the `zeroth-order method').  We then extend the physical model to account for realistic path-length differences and present an iterative numerical method that is able to correct analytically derived TMs to satisfy this extended model (the `first-order method').  Neither method requires that the TM be unitary, only that it is invertible, meaning they could be applied to lossy fibres or general scattering media. Following TM reconstruction, we then show how imaging can be performed at a fourth wavelength.

Next, we computationally demonstrate the first-order method using simulated 32$\times$32 non-unitary fibre matrices with noise.  We then demonstrate the recovery algorithm using measured MCF TMs ($1648 \times 1648$) and realistically simulated reflector stacks, showing the method can be scaled up effectively. Finally, we demonstrate TM recovery using a multi-wavelength set of TMs measured from a step-index MMF validating both that the method works for real MMF and that the first-order physical model used to predict changes in TM with wavelength is physically realistic. Our findings underpin a compelling new approach for enabling ultra-thin flexible lensless imaging via optical fibres.

\section{Theory}
\subsection{Formulation of the physical model}
\label{subsec:physicalModel}
\begin{figure}[htbp]
	\centering
	\includegraphics[width=\linewidth]{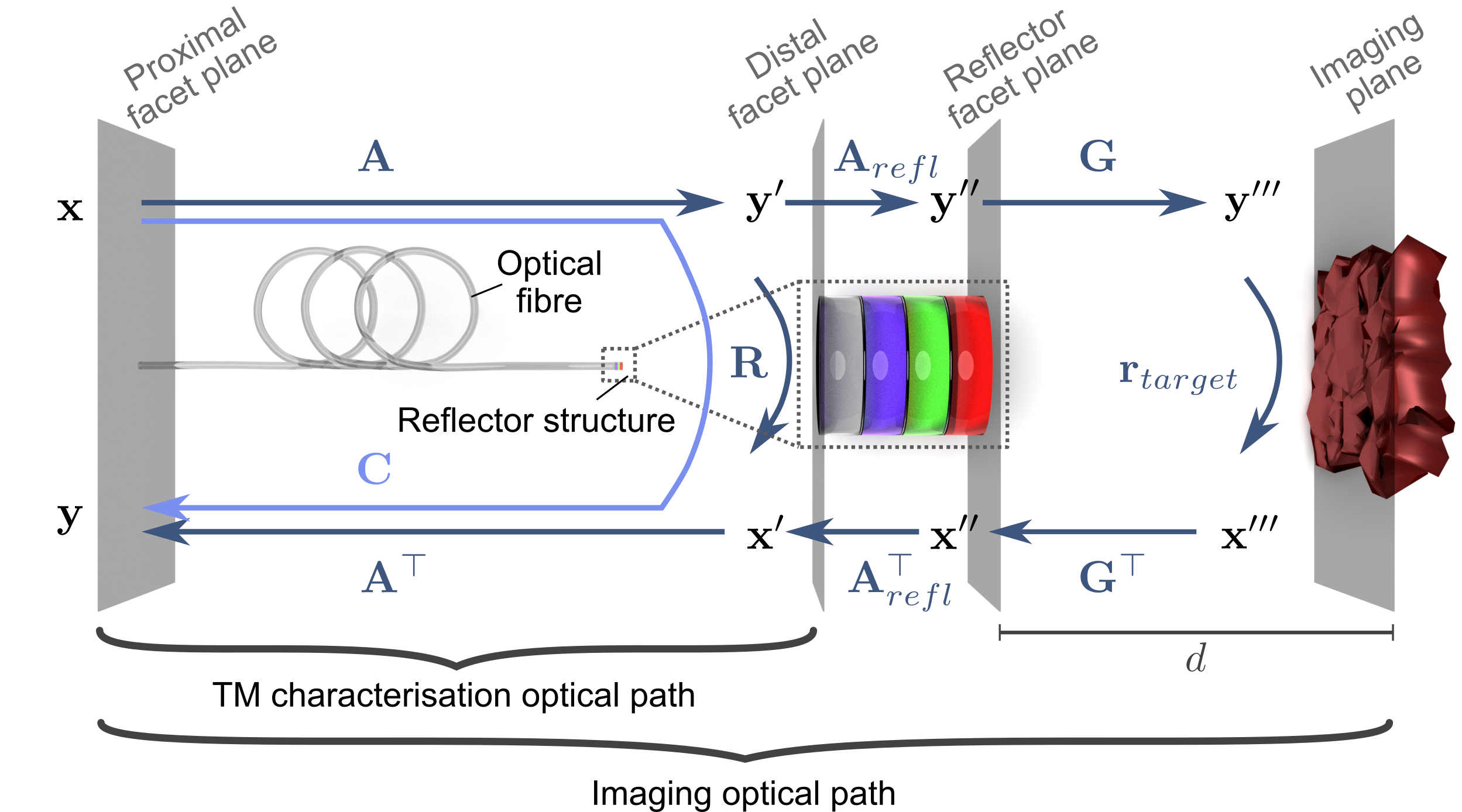}
	\caption{\textbf{Physical model used for fibre TM characterisation in reflection mode and for imaging.} The proximal facet of the fibre is nearest to the imaging system (i.e. outside of the subject) while the distal facet is nearest the object being imaged (i.e. inside the subject). Vectors at the distal facet are denoted using the prime symbol ($'$), vectors at the exit facet of the reflector stack are denoted with double prime ($''$), and vectors at the imaging plane are denoted with triple prime ($'''$). $\mathbf{A}$ represents the fibre TM, $\mathbf{R}$ the PM of the reflector stack, $\mathbf{C}$ the round-trip RM (down the fibre, off the reflector, back up the fibre), $\mathbf{A}_{refl}$ the TM of the reflector stack, $\mathbf{G}$ a free-space propagation operator (parameterised by the distance, $d$, between the reflector structure and the imaging plane), and $\mathbf{r}_{target}$ the reflectance profile of the sample.}
	\label{fig:physicalModel}
\end{figure}

The physical model used for reference in this paper is shown in Figure \ref{fig:physicalModel}. We consider a well-established model where a field containing $M$ pixels, each complex-valued to encode amplitude and phase, in two orthogonal polarisations is recorded. The field at some input plane (e.g. the proximal facet), is represented by a vector $\mathbf{x} \in \mathbb{C}^{2M}$, and the field at an output plane (e.g. the distal facet) is represented by a vector $\mathbf{y}' \in \mathbb{C}^{2M}$. In the forward propagation direction, these vectors are related by the fibre TM, $\mathbf{A} \in \mathbb{C}^{2M \times 2M}$:

\begin{equation}
\label{eq:basicForward}
\mathbf{y}' = \mathbf{A} \mathbf{x}
\end{equation}

This equation can be derived by discretisation of the linear propagation relationships between the two planes \cite{Rotter2017,Fan2005}. For example, a continuous optical field might be sampled using delta functions at $M$ spatial locations in two orthogonal polarisation states using a Jones vector formalism \cite{Goldstein2010, Gataric2018}. The complex-valued samples, representing amplitude and phase across two polarisations, can then be ordered to produce the vector $\mathbf{x}$ with $2M$ elements \cite{Gordon2018}.

Similarly, if we consider a field, $\mathbf{x}' \in \mathbb{C}^{2M}$, at the distal facet propagating in the reverse direction to become a field, $\mathbf{y} \in \mathbb{C}^{2M}$, at the proximal facet, these are related by the transpose of the fibre TM \cite{Gu2015}:

\begin{equation}
\label{eq:basicBackward}
\mathbf{y} = \mathbf{A}^{\top} \mathbf{x}'
\end{equation}

We then consider the impact of adding a reflector at the distal facet of the fibre. In general, the reflector is considered to be spatially heterogeneous in terms of its localised Jones reflection matrices: there may be uncorrelated Jones matrices describing reflections at each spatial point.  Further, if the reflector is offset from the fibre, light may couple between spatial positions due to diffraction.  This behaviour is linear and so is represented by a partial reflector matrix (PM), $\mathbf{R} \in \mathbb{C}^{2M \times 2M}$, that relates $\mathbf{y}'$ and $\mathbf{x}'$ (see Figure \ref{fig:physicalModel}):

\begin{equation}
\label{eq:reflector}
\mathbf{x}' = \mathbf{R} \mathbf{y}'
\end{equation}

Next, combining Equations \ref{eq:basicForward}, \ref{eq:basicBackward} and \ref{eq:reflector} we can determine the optical field exiting the proximal facet in the reverse direction:

\begin{equation}
\label{eq:reflectionMatrix}
\mathbf{y} = \mathbf{A}^{\top} \mathbf{R} \mathbf{A} \mathbf{x} = \mathbf{C} \mathbf{x}
\end{equation}

\noindent where $\mathbf{C} = \mathbf{A}^{\top} \mathbf{R} \mathbf{A} \in \mathbb{C}^{2M \times 2M}$ is termed a `reflection matrix' (RM). Physically speaking, $\mathbf{C}$ represents light taking a complete round-trip (or double-pass): forwards down the fibre, off a given reflector and back up the fibre, as illustrated in Figure \ref{fig:physicalModel}. When $\mathbf{C}$ is measured in practice, the measurements may lead to a non-square approximation to $\mathbf{C}$, denoted as $\tilde{\mathbf{C}}$, hence $\tilde{\mathbf{C}}$ first needs to be downsampled into a square matrix $\tilde{\tilde{\mathbf{C}}}$, which is then used as a surrogate for $\mathbf{C}$ in what follows (see Appendix \ref{subsec:downsamplingAppendix}).

As our first objective is to estimate the TM of the fibre $\mathbf{A}$ we would ideally use pairs of measured vectors, $(\mathbf{x},\mathbf{y}')$ through standard a calibration procedure \cite{Gordon2018,Gataric2018,Carpenter2014}.  Since we now only have access to the proximal facet of the fibre, we instead need to estimate the RM, $\mathbf{C} = \mathbf{A}^{\top} \mathbf{R} \mathbf{A}$, based on the measured vectors $(\mathbf{x},\mathbf{y})$. Given that we can characterise $\mathbf{R}$ in advance (Appendix \ref{subsubsec:charBeforeUseAppendix}) and it will remain fixed throughout use, the goal is then to recover $\mathbf{A}$ based on measurements of $\mathbf{C}$ and prior characterisations of $\mathbf{R}$, however, with only this information it is not in general possible to recover $\mathbf{A}$ unless the fibre TM is unitary \cite{Gu2015,Gu2017}.  We therefore propose a novel approach that uses several different reflectors, with PMs $\mathbf{R}_n,~n=1..Q$, and show that this enables unambiguous recovery of $\mathbf{A}$ in the more general non-unitary case.

For experimental simplicity, it would be desirable to use as few different reflectors, $\mathbf{R}_n,~n=1..Q$, and associated RMs, $\mathbf{C}_n,~n=1.. Q$, as possible for TM characterisation and to switch between these without any physical moving parts, for example, by modulating wavelength. The case with a single reflector reduces to the method presented by Gu et al. \cite{Gu2015} and does not produce unique solutions for non-unitary TMs, $\mathbf{A}$.  Therefore, we need $Q \geq 2$ reflectors. 

We will consider in what follows the case of three reflectors, i.e. $Q=3$. For TM characterisation (with reference to the appropriate optical path in Figure \ref{fig:physicalModel}) there will be three associated PMs $\mathbf{R}_n,~n=1..3$.  The TMs associated with each reflector during characterisation, $\mathbf{A}_n, n=1..3$, may in general be different, for example if wavelength modulation is used to activate different reflectors.  We can then record 3 associated RMs, $\mathbf{C}_n,~n=1..3$ measured for example using the process of Appendix \ref{subsubsec:instTMAppendix}.  Based on Equation \ref{eq:reflectionMatrix}, we express these RMs in terms of the PMs $\mathbf{R}_n,~n=1..3$ and associated TMs $\mathbf{A}_n,~n=1..3$ as:

\begin{equation}
\label{eq:C1}
\mathbf{C}_1 = \mathbf{A}_1^{\top} \mathbf{R}_1 \mathbf{A}_1
\end{equation}

\begin{equation}
\label{eq:C2}
\mathbf{C}_2 = \mathbf{A}_2^{\top} \mathbf{R}_2 \mathbf{A}_2
\end{equation}

\begin{equation}
\label{eq:C3}
\mathbf{C}_3 = \mathbf{A}_3^{\top} \mathbf{R}_3 \mathbf{A}_3
\end{equation}

In order to solve these equations for $\mathbf{A}_n, n=1..3$ we must assume a known relationship between these TMs, for example a wavelength-dependent phase shift, and must also have measured $\mathbf{R}_1$, $\mathbf{R}_1$ and $\mathbf{R}_1$ \emph{a priori} using a characterisation process such as that in Appendix \ref{subsubsec:charBeforeUseAppendix}.

Following TM recovery, the final objective is to performing imaging. This involves using an extended optical path (Figure \ref{fig:physicalModel}) in which light propagates down the fibre via a possibly different fourth TM, $\mathbf{A}_4$, is partially reflected off the reflector via a PM, $\mathbf{R}_4$, and is partially transmitted through the reflector via a TM, $\mathbf{A}_{refl}$.  $\mathbf{R}_4$ and $\mathbf{A}_refl$ can be measured \emph{a priori} using a characterisation process such as that in Appendix \ref{subsubsec:charBeforeUseAppendix}. Following possible further propagation and reflection off the sample, the returned light at the proximal facet is used for image reconstruction. 

Again, a known physical relationship is required to recover $\mathbf{A}_4$ from $\mathbf{A}_1$, $\mathbf{A}_2$ or $\mathbf{A}_3$ and hence reconstruct images. We therefore consider two simple models to approximate this relationship between the calibration and imaging TMs and enable image recovery: a `zeroth-order model' and a `first-order model'. 

\subsection{Zeroth-order model}
\label{subsec:zeroordermodel}
The zeroth-order model assumes that the TMs under the different reflectors are the same, i.e.:

\begin{equation}
\label{eq:zerothOrderAssumption}
\mathbf{A} = \mathbf{A}_1 = \mathbf{A}_2 = \mathbf{A}_3 = \mathbf{A}_4
\end{equation}

This could correspond to physically switching reflectors without disturbing the fibre TM.  Alternatively, if modulating wavelength to switch between reflectors, this corresponds to the assumption that the fibre TM does not change with wavelength. Using this model, the TM can be recovered in a computationally straightforward way relying largely on analytical steps (Appendix \ref{subsec:zeroordermodelapp}), providing a number of key insights.  First, if the number of reflectors is $n \geq 3$ then analytical solutions for the TMs can be recovered.  Numerical and analytical investigations for the case with $n=2$ reflectors were inconclusive.  Second, the eigenvalues of each PM, $R_n, n=1..3$ must be distinct for unambiguous TM recovery, in agreement with previous findings \cite{Gu2015}. This means that light must be coupled between different modes and polarisations and so, for example, a conventional partial mirror reflector would not work because it preserves polarisation states leading to repeated matrix eigenvalues.  This insight influences the potential design of any reflector.

Though this model may be used if reflectors are physically switched, for the wavelength modulation approach, Equation \ref{eq:zerothOrderAssumption} is physically unrealistic because it neglects wavelength-induced phase-shifts that are significant even over small physical distances (see Appendix \ref{sec:bwCalcAppendix}).  However, even if these phase-shifts are neglected the recovered TMs still provide an approximate starting point for more complex models relating TMs at different wavelengths.

\subsection{First-order model}
\label{subsec:firstordermodel}
Considering again the idea of switching reflectors by modulating wavelength, a more physically accurate relationship between TMs is constructed by considering linear phase shift as wavelength is varied.  In general the relationship between fibre TMs at different wavelengths is complex \cite{Redding2013a} so this linear approximation is limited to be within the `spectral bandwidth' of the fibre. We can then use a coupled-mode theory treatment of optical fibres \cite{Yariv1973,Huang1994}.  This perturbational approach models a length of MMF as a sequence of infinitesimal segments that introduce field coupling.  The differential change in field in each mode can be modelled by a vector, $d\mathbf{m} \in \mathbb{C}^{2M \times 1}$, such that:

\begin{equation}
d\mathbf{m} = d\mathbf{A} \mathbf{m}
\end{equation}

\noindent where $d\mathbf{A} \in \mathbb{C}^{2M \times 2M}$ represents an infinitesimal coupling matrix, and $\mathbf{m} \in \mathbb{C}^{2M \times 1}$ represents the field in each of $M$ spatial modes over 2 orthogonal polarisations.  This matrix equation represents a system of $2M$ linear differential equations. The solution to these equations for any arbitrary input set of modes, $\mathbf{x}$, after travelling a distance $\ell$ along the central axis of the fibre is given by a matrix exponential \cite{Yariv1973,Loterie2017}:

\begin{equation}
\mathbf{y} = e^{d\mathbf{A}\ell} \mathbf{x}
\end{equation}

Equivalently, this model can be thought of as the Lie algebra $d\mathbf{A}$ constructing the Lie group of invertible matrices $e^{d\mathbf{A}\ell} \in \mathbb{C}^{2M \times 2M}$ \cite{Dragt1982}.  We can estimate $d\mathbf{A}$ from the TM at characterisation wavelength $\lambda_1$, $\mathbf{A}_1$ as:

\begin{equation}
\label{eq:dAfromA1}
d\mathbf{A} = \frac{\log \mathbf{A}_1}{\ell_1}
\end{equation}

\noindent where $\log \mathbf{A}_1$ represents a matrix logarithm and $\ell_1$ is the optical path length along the central fibre axis. In general, matrix logarithms of complex matrices produce degenerate solutions, analogous to logarithms of complex numbers.  This means that in estimating $d\mathbf{A}$ each eigenvalue, $\gamma$, has an equivalence class of degenerate forms: $\gamma + i 2\pi n / \ell_1$, $n \in \mathbb{Z}$.  However, under the bandwidth-limited scenario presented here, the change in wavelength results in an overall relative phase shift between any two matrix elements of $<2\pi$ (Appendix \ref{sec:bwCalcAppendix}) and so therefore $n=0$ is the appropriate choice.

If we assume constant dispersion within the relatively small ($<10$nm) spectral bandwidth considered, a reasonable assumption for typical glasses \cite{Payne1975}, we can find the equivalent optical thicknesses at the other characterisation wavelengths ($\lambda_2$, $\lambda_3$) and imaging wavelength ($\lambda_4$) relative to the first characterisation wavelength ($\lambda_1$):

\begin{equation}
\ell_2 = \frac{\lambda_1}{\lambda_2} \ell_1
\end{equation}

\begin{equation}
\ell_3 = \frac{\lambda_1}{\lambda_3} \ell_1
\end{equation}

\begin{equation}
\ell_4 = \frac{\lambda_1}{\lambda_4} \ell_1
\end{equation}

Rearranging and expressing in terms of $\mathbf{A}_1$ we can write

\begin{equation}
\label{eq:firstOrderA2}
\mathbf{A}_2 = e^{d\mathbf{A} \ell_2} = e^{d\mathbf{A} \frac{\lambda_1}{\lambda_2} \ell_1} = e^{\frac{\lambda_1}{\lambda_2} \log \mathbf{A}_1}
\end{equation}

\begin{equation}
\label{eq:firstOrderA3}
\mathbf{A}_3 = e^{\frac{\lambda_1}{\lambda_3} \log \mathbf{A}_1}
\end{equation}

\begin{equation}
\label{eq:firstOrderA4}
\mathbf{A}_4 = e^{\frac{\lambda_1}{\lambda_4} \log \mathbf{A}_1}
\end{equation}

A key requirement for this model is that the wavelength sweep range is kept within the spectral bandwidth of the fibre, i.e. the bandwidth within which far-field speckle patterns from a fibre remain correlated \cite{Mosk2012} or equivalently over which the eigenmodes of the fibre remain approximately the same. We show in Appendix \ref{sec:bwCalcAppendix} that this implies the maximum change in path length for any mode is $< 2 \pi$. 

The eigendecomposition of $d\mathbf{A}$ is:

\begin{equation}
d\mathbf{A} = \mathbf{Q}
\left[\begin{array}{c c c} 
\Lambda_1 & \cdots & 0 \\
\vdots & \ddots & \vdots \\
0 & \cdots & \Lambda_{2M}
\end{array}\right]
\mathbf{Q}^{-1}
\end{equation}

\noindent where $\mathbf{Q} \in \mathbb{C}^{2M \times 2M}$ is a matrix whose columns, $\mathbf{q}_m$ with $m=1.. 2M$, are the eigenvectors of $d\mathbf{A}$ and $\Lambda_m$ with $m=1.. 2M$ are the corresponding eigenvalues. The eigendecomposition of $\mathbf{A}_n$ with $n=2,3,4$ is therefore \cite{Hall2015}:

\begin{equation}
\mathbf{A}_n = e^{\frac{\lambda_1}{\lambda_n} \log \mathbf{A}_1} = \mathbf{Q}
\left[\begin{array}{c c c} 
e^{\frac{\lambda_1}{\lambda_n} \Lambda_1} & \cdots & 0 \\
\vdots & \ddots & \vdots \\
0 & \cdots & e^{\frac{\lambda_1}{\lambda_n} \Lambda_{2M}}
\end{array}\right]
\mathbf{Q}^{-1}
\end{equation}

In general, matrices constructed using matrix exponentials (as in Equations \ref{eq:firstOrderA2} -- \ref{eq:firstOrderA4}) share the same eigenvectors but the corresponding eigenvalues are related by an exponential \cite{Hall2015}. Our model therefore assumes that the eigenvectors of the fibre do not change appreciably over the selected bandwidth, which is consistent with previous experimental work \cite{Carpenter2015}.  This is in contrast to the zeroth-order model that requires the additional assumption that eigenvalues stay constant within the spectral bandwidth.

It is not straightforward to solve Equation \ref{eq:C1} to Equation \ref{eq:C3} analytically to obtain $\mathbf{A}_1$, $\mathbf{A}_2$, $\mathbf{A}_3$ or $\mathbf{A}_4$ when they are related by matrix exponentials as in Equations \ref{eq:firstOrderA2} -- \ref{eq:firstOrderA4}.  Therefore, we present an optimisation-based approach.  We begin by using the zeroth-order approximation to obtain an approximate solution for $\mathbf{A}_1$, termed $\mathbf{\hat{A}}_1$.  This is achieved by performing an optimisation based on Equation \ref{eq:C1}:

\begin{equation}
\label{eq:opt1}
\mathbf{\hat{A}}_1' = \arg \min_{\mathbf{A}} \left\lVert \mathbf{A}^{\top} \mathbf{R}_1 \mathbf{A} - \mathbf{C}_1 \right\rVert_F
\end{equation}

\noindent where $\left \lVert .. \right \rVert_F$ indicates the Frobenius norm. The solution to this optimization problem, $\mathbf{\hat{A}}'_1$, may be found using an iterative gradient descent solver initialised by the zeroth-order approximation.  We know that because the true $\mathbf{A}_1$ may be non-unitary, this equation may not have a unique solution \cite{Gu2015}.  We therefore use the result of Equation \ref{eq:opt1}, $\mathbf{\hat{A}}_1'$, to generate an estimate of $\mathbf{A}_2$, as $\mathbf{\hat{A}}_2 = e^{\frac{\lambda_1}{\lambda_2}\log(\mathbf{\hat{A}}_1')}$. This is then used as the starting point for a second optimisation:

\begin{equation}
\label{eq:opt2}
\mathbf{\hat{A}}_2' = \arg \min_{\mathbf{A}} \left\lVert \mathbf{A}^{\top} \mathbf{R}_2 \mathbf{A} - \mathbf{C}_2 \right\rVert_F
\end{equation}

Finally, $\mathbf{\hat{A}}_3 = e^{\frac{\lambda_2}{\lambda_3} \log(\mathbf{\hat{A}}_2')}$  is used as the starting point for a third optimisation:

\begin{equation}
\label{eq:opt3}
\mathbf{\hat{A}}_3' = \arg \min_{\mathbf{A}} \left\lVert \mathbf{A}^{\top} \mathbf{R}_3 \mathbf{A} - \mathbf{C}_3 \right\rVert_F
\end{equation}

\begin{figure}[htpb!]
	\centering
	\includegraphics[height=0.9\textheight]{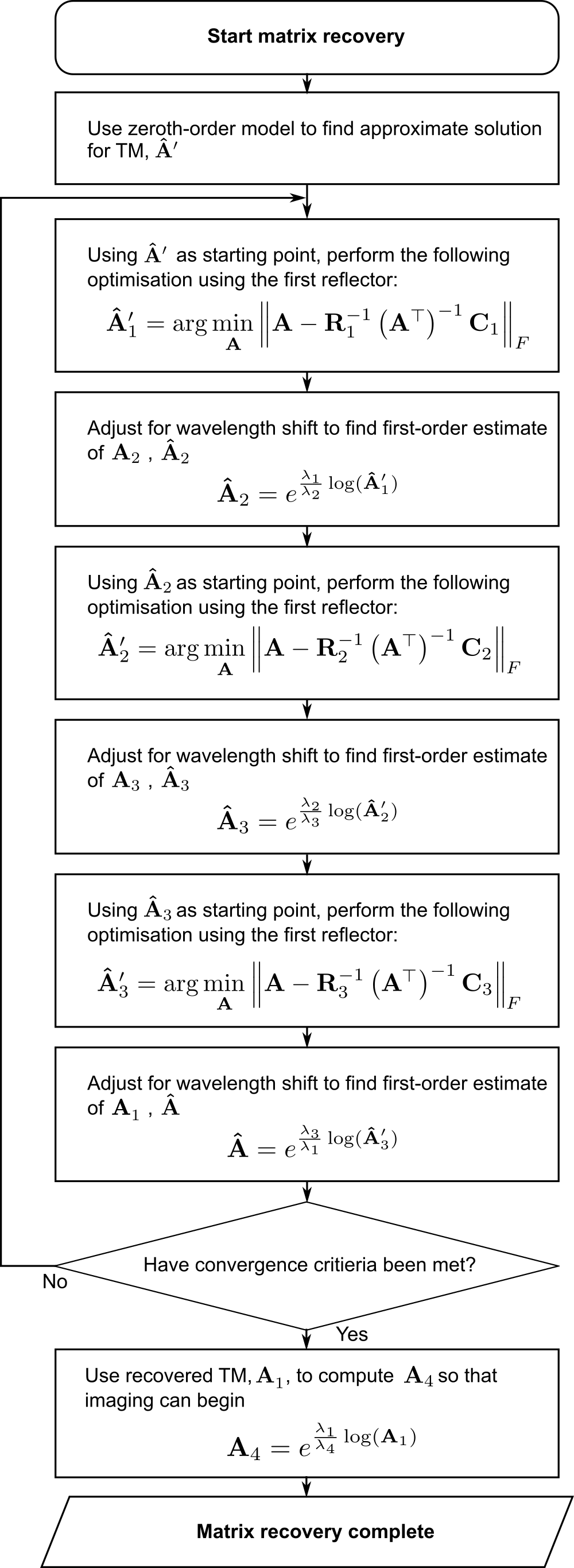}
	\caption{\textbf{Flowchart detailing the iterative optimisation routine to estimate the TM required for imaging, $\mathbf{A}_4$, using the first-order model.}}
	\label{fig:firstOrderFlowchart}
\end{figure}

To complete the iteration, $\mathbf{\hat{A}}_1 = e^{\frac{\lambda_3}{\lambda_1} \log(\mathbf{\hat{A}}_3')}$ is used as a starting point for Equation \ref{eq:opt1}, and the process is repeated.  For numerous simulated and experimentally measured TMs we observe that $\mathbf{\hat{A}}_1$ converges to the true value of $\mathbf{A}_1$, thereby enabling calculation of $\mathbf{A}_4$, required for image reconstruction.  This process is summarised in the flow-chart of Figure \ref{fig:firstOrderFlowchart}. In this work, for simulated TMs the convergence criteria is a fixed number of iterations (both 300 and 500 are examined) while for experimentally measured TMs the convergence criteria is an error gradient (terminate when change in error is $<$1\% of value at first iteration).

Each of Equation \ref{eq:opt1}--\ref{eq:opt3} is in itself degenerate due to the quadratic matrix form $\mathbf{A}^{\top} \mathbf{R} \mathbf{A}$ (see Section \ref{subsec:physicalModel}).  However, this iterative process may avoid local minima because such degeneracies are not typically the same for all three equations. The exception to this is a global sign error: $\pm \mathbf{A}_n$ with $n=1.. 3$, is optimal for each respective equation.  However, this degeneracy corresponds to a global phase-shift (i.e. a constant factor) so can be neglected for almost all imaging modalities.

Local optima can still arise when the optimisation of Equation \ref{eq:opt3} exactly reverses the optimisations of Equations \ref{eq:opt1} and \ref{eq:opt2}. It is observed empirically that using the zeroth-order approximation as a starting point tends to avoid this and produces reliable convergence to the global minimum.

The algorithm presented here is observed to converge reliably to the correct solution and has the computational advantage that the relatively expensive matrix logarithm operation is performed only between minimisations, as opposed to in each objective function.  It is therefore used for the remainder of this paper.

\subsection{Imaging}
\label{subsec:imaging}
Considering the case in which reflectors are switched by modulating wavelength and the first-order model is used to reconstruct the fibre TM at the imaging wavelength ($\lambda_4$), $\mathbf{A}_4$, the next step is to reconstruct an image of a sample using data recorded at the imaging wavelength. With reference to the physical model (Figure \ref{fig:physicalModel}) this requires prior measurement of the TM and PM of the reflector, $\mathbf{A}_{refl}$ and $\mathbf{R}_4$ respectively, at the imaging wavelength, $\lambda_4$, via a calibration process (e.g. Appendix \ref{subsubsec:charBeforeUseAppendix}). These quantities are assumed to stay constant throughout operation. The imaging process is summarised in the flow chart of Figure \ref{fig:sampleImaging}.  An experimental system in which such a workflow could be realised is described in Appendix \ref{subsec:expSetupAppendix}, where a spatial light modulator (SLM) is used to project optical fields onto the proximal fibre facet. 

\begin{figure}[htpb]
	\centering
	\includegraphics[height=0.8\textheight]{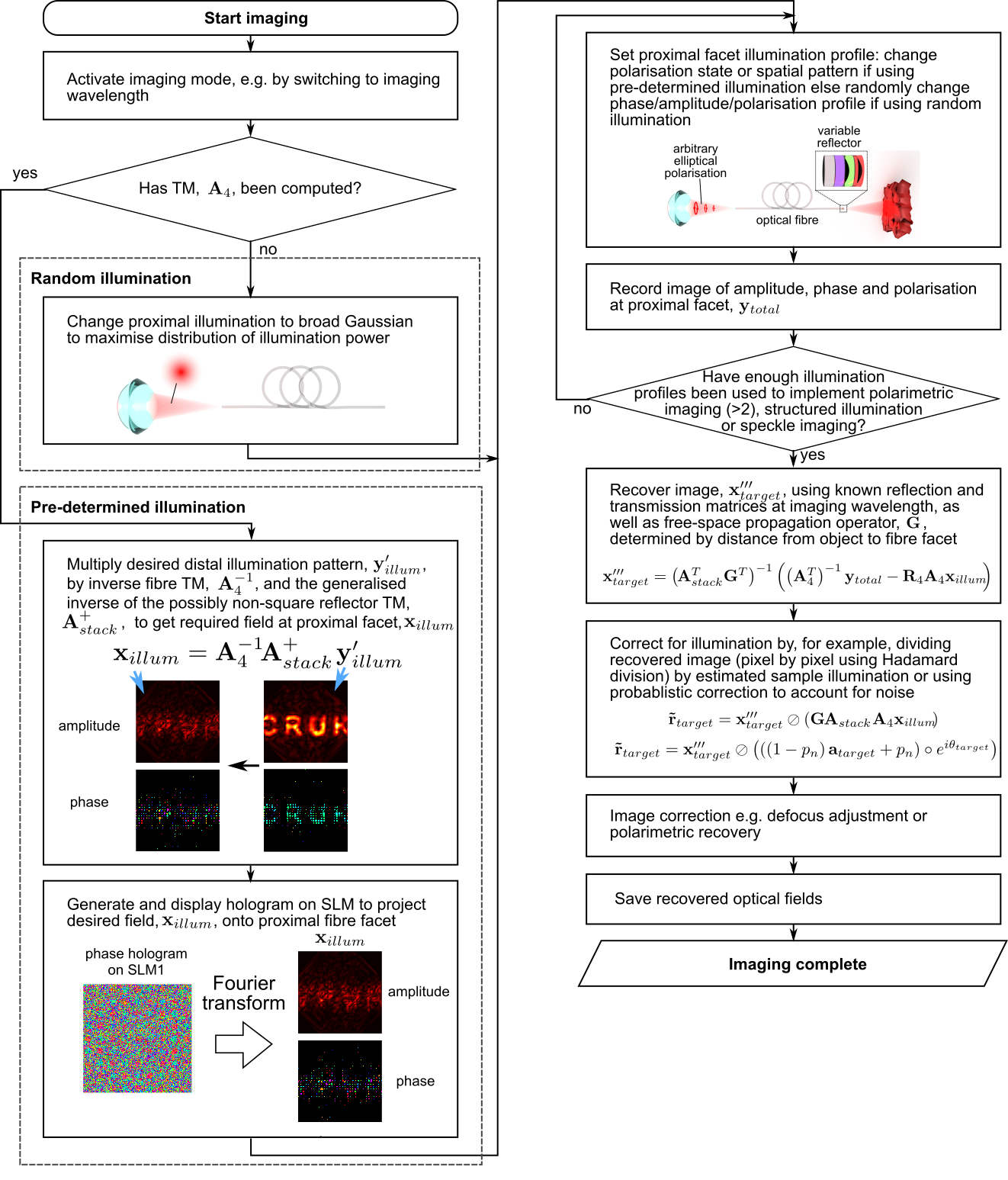}
	\caption{\textbf{Flowchart detailing how an image of a sample is recorded.}}
	\label{fig:sampleImaging}
\end{figure}

There are two approaches to illuminating the sample via the imaging fibre. The first is to use the measured TM at the imaging wavelength, $\mathbf{A}_4$, to produce a known distal illumination profile. This produces controllable illumination profiles that could enable, for example, various structured illumination techniques but assumes that the TM reconstruction process is complete before imaging. In cases where computing the TM introduces an undesirable time delay, a known proximal illumination profile could used to produce a semi-random distal illumination profile that can later be corrected by averaging over several random illuminations, a technique known as speckle averaging \cite{Choi2012}, or by using knowledge of $\mathbf{A}_4$ reconstructed offline.

Considering our physical model (Figure \ref{fig:physicalModel}), we project a random or pre-calculated illumination vector, $\mathbf{x}_{illum}$, onto the proximal facet of the fibre, giving at the distal facet:

\begin{equation}
\mathbf{y}_{illum}' = \mathbf{A}_4 \mathbf{x}_{illum}
\end{equation}

Next, the light passes through the full reflector structure, which in the imaging configuration must be partially transmissive. In general $\mathbf{A}_{refl} \in \mathbb{C}^{2N \times 2M}$ with $N \geq M$, recalling that $\mathbf{A}_4 \in \mathbb{C}^{2M \times 2M}$.  This allows for oversampled illumination fields.  The field exiting the reflector stack is then:

\begin{equation}
\mathbf{y}''_{illum} =  \mathbf{A}_{refl} \mathbf{A}_4 \mathbf{x}_{illum}
\end{equation}

\noindent with $\mathbf{y}''_{illum}  \in \mathbb{C}^{2N}$. This can be propagated by some linear operator, $\mathbf{G} \in \mathbb{C}^{2N \times 2N}$, through free-space (e.g. Fresnel or Fraunhofer propagation) before reaching the target where the field will be:

\begin{equation}
\label{eq:xTarget}
\mathbf{y}_{illum}''' =  \mathbf{G} \mathbf{A}_{refl} \mathbf{A}_4 \mathbf{x}_{illum}
\end{equation}

In general $\mathbf{G}$ is parameterised by the distance, $d$, between the subject and the distal surface of the reflector structure (Figure \ref{fig:physicalModel}.  In general, $d$ is not known \emph{a priori} and so must be estimated during operation, for example using an automatic focus detection algorithm such as the Brenner algorithm \cite{Brenner1976, Gordon2018}.

For simplicity we will assume the target is a reflective surface, though a volume scattering target could be characterised using several known illumination profiles in sequence as in spatial frequency domain imaging \cite{Cuccia2009}. We therefore let the target be represented by vector, $\mathbf{r}_{target} \in \mathbb{C}^{2N}$. An element-wise (Hadamard) product with the illumination gives:

\begin{equation}
\mathbf{x}_{target}''' =  \mathbf{y}_{illum}''' \circ \mathbf{r}_{target}
\end{equation}

This in turn, produces a field at the distal fibre facet of:

\begin{equation}
\label{eq:imtarget1}
\mathbf{x}_{target}' =  \mathbf{A}_{refl}^{\top} \mathbf{G}^{\top} \mathbf{x}_{target}'''
\end{equation}

Next, we consider the illumination light that is reflected back from the reflector stack:

\begin{equation}
\label{eq:imrefl1}
\mathbf{x}_{refl}' = \mathbf{R}_4 \mathbf{A}_4 \mathbf{x}_{illum}
\end{equation}

We sum the two fields and propagate back through the fibre giving:

\begin{equation}
\label{eq:ytotalEq}
\mathbf{y}_{total} = \mathbf{A}_4^{\top}(\mathbf{x}_{target}'  + \mathbf{x}_{refl}')
\end{equation}

The goal is to recover $\mathbf{r}_{target}$ from $\mathbf{y}_{total}$, the raw measured data.  To do this, we first recover $\mathbf{x}_{target}'''$ by rearrangement of Equation \ref{eq:ytotalEq} and substitution of Equations \ref{eq:imtarget1} and \ref{eq:imrefl1}:

\begin{equation*}
\left(\mathbf{A}_4^{\top}\right)^{-1}\mathbf{y}_{total} = (\mathbf{x}_{target}' + \mathbf{x}_{refl}')
\end{equation*}

\begin{equation*}
\mathbf{x}_{target}' = \left(\mathbf{A}_4^{\top}\right)^{-1}\mathbf{y}_{total} -  \mathbf{R}_4 \mathbf{A}_4 \mathbf{x}_{illum}
\end{equation*}

\begin{equation}
\label{eq:imRecovery}
\mathbf{x}_{target}''' = \left(\mathbf{A}_{refl}^{\top} \mathbf{G}^{\top} \right)^{+} \left(\left(\mathbf{A}_4^{\top}\right)^{-1}\mathbf{y}_{total} -  \mathbf{R}_4 \mathbf{A}_4 \mathbf{x}_{illum}\right)
\end{equation}

\noindent where $(..)^+$ represents a generalised inverse because $\mathbf{A}_{refl}$ may be non-square. Quantities on the RHS of Equation \ref{eq:imRecovery} are known prior to imaging, except for $\mathbf{y}_{total}$ which is recorded during imaging and possibly $\mathbf{G}$, which may need to be estimated as described above. Therefore, we could naively recover an estimate of $\mathbf{r}_{target}$, which we term $\mathbf{\tilde{r}}_{target}$, from $\mathbf{x}_{target}'''$ through an element-wise (Hadamard) division:

\begin{equation*}
\mathbf{\tilde{r}}_{target} = \mathbf{x}_{target}''' \oslash \mathbf{y}_{illum}'''
\end{equation*}

\begin{equation*}
\mathbf{\tilde{r}}_{target} = \mathbf{x}_{target}''' \oslash \left( \mathbf{G} \mathbf{A}_{refl} \mathbf{A}_4 \mathbf{x}_{illum} \right)
\end{equation*}

However, this approach may erroneously amplify noise in areas where illumination power, $|\mathbf{y}_{illum}'''|^2$, is low producing speckle-like artefacts in the recovered $\mathbf{\tilde{r}}_{target}$.  To compensate for these artefacts we apply a probabilistic approach.  Using the model derived in Section \ref{subsec:noiseModel}, the noise at pixel $(x,y)$ is considered circularly-symmetric complex Gaussian distributed with standard deviation $\sigma$. The noise power at that pixel, termed $\mathrm{n}(x,y)$, is therefore drawn from a $\chi^2$ distribution with 1 degree of freedom:

\begin{equation*}
\frac{\mathrm{n}(x,y)}{\sigma^2} \sim \chi^2 (1)
\end{equation*}

Using an empirically measured value for $\sigma$ we use the $\chi^2$ cumulative distribution function, $\chi^2_{CDF}$, to estimate for each pixel a probability that it is generated by noise, $p_n(x,y)$:

\begin{equation*}
p_n(x,y) = \chi^2_{CDF}\left(\frac{||\mathbf{x}_{target}'''(x,y)||^2}{\sigma^2}\right)
\end{equation*}

We can order $p_n(x,y)$ across all $(x,y)$ into a vector denoted as $\mathbf{p}_n$. Separating the illumination into amplitude and phase parts as:

\begin{equation*}
\mathbf{a}_{illum} = |\mathbf{y}_{illum}'''|
\end{equation*}

\begin{equation*}
\mathbf{\theta}_{illum} = \arg (\mathbf{y}_{illum}''')
\end{equation*}

\noindent we then use element-wise division by a probabilistically weighted term as follows:

\begin{equation}
\label{eq:illumCorrectionProb}
\mathbf{\tilde{r}}_{target} = \mathbf{x}_{illum}''' \oslash \left(\left(\left(1-\mathbf{p}_{n}\right)\circ \mathbf{a}_{illum} + \mathbf{p}_{n}\right) \circ e^{i\mathbf{\theta}_{target}}\right)
\end{equation}

With this approach, pixels with smaller illumination amplitudes tend only to modify the phase of $\mathbf{x}_{illum}'''$ when correcting the speckle-like artefacts during recovery of $\mathbf{\tilde{r}}_{target}$.  Pixels with larger illumination amplitudes, more likely to contain useful information, experience both amplitude and phase correction.  In this way, the speckle-like artefacts in $\mathbf{\tilde{r}}_{target}$ are reduced. Areas of low illumination will still have poor signal-to-noise ratio but this could be corrected by using a more uniform pre-determined illumination profile or, again, by implementing speckle averaging \cite{Choi2012}.

\section{Methods}
\subsection{Implementation of reflector}
\label{subsec:implementingReflectors}
\begin{figure}[htbp]
	\centering
	\includegraphics[width=0.85\linewidth]{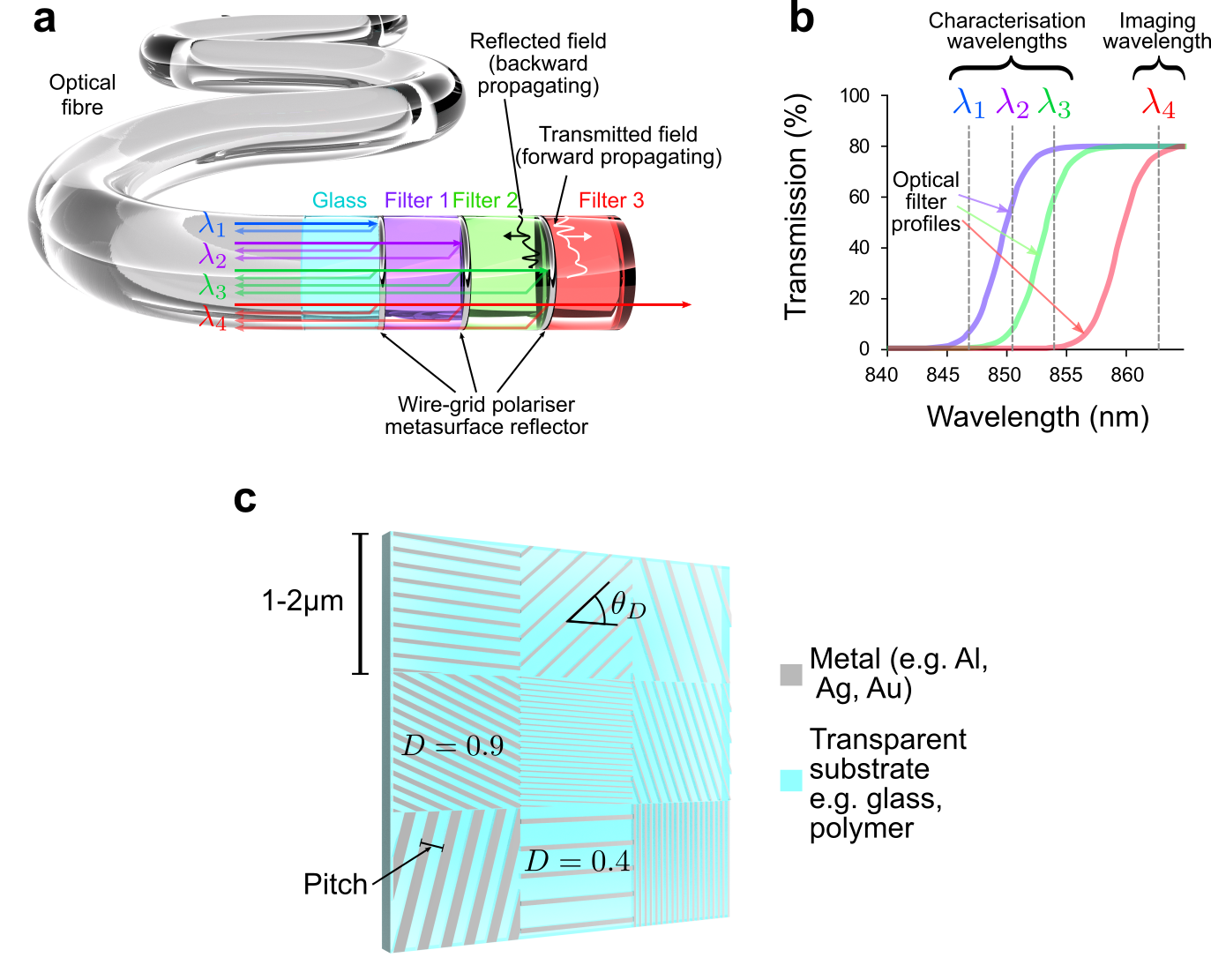}
	\caption{\textbf{Design of a wavelength-dependent reflector stack}:  a) Proposed stack structure, attached to distal facet of optical fibre, comprising layers of glass and long-pass optical filters with wire-grid polariser metasurfaces sandwiched between.  Light paths are shown for the 4 different wavelenths used: the characterisation wavelengths $\lambda_1$, $\lambda_2$ and $\lambda_3$ each producing 3 reflector matrices, $\mathbf{R}_n~n=1..3$; and the imaging wavelength $\lambda_4$ that passes the whole way through the stack. The stack behaviour can be understood by considering optical fields exiting the fibre, propagating through each layer via diffraction and being partially reflected at the metasurfaces generating backward propagating wavefronts.  The sum of all backward propagating components represents the field coupled back into the fibre. b) Spectral transmission profiles of the 3 long-pass optical filters with example imaging and characterisation wavelengths labelled. c) Example design of wire-grid polariser metasurface comprising an array of diattenuating cells of size 1--2$\mu$m containing nanowire gratings with varying pitch (50--200nm), duty cycle (25\%--75\%) and orientation (0--2$\pi$). The pitch and duty cycle determine the linear diattenuation of each cell, $D$, and the orientation determines the diattenuation axis, $\theta_D$.  These are used to form a Jones matrix for each cell that determines its optical transmission and reflection.}
	\label{fig:reflectorDesign}
\end{figure}

The switchable reflector structure described in Section \ref{subsec:physicalModel} could be implemented in a number of ways. Reflectors could be switched using a physical mechanism, e.g by translating or rotating a large distal plate relative to the facet.  Though this would enable use of the computationally simpler zeroth-order reconstruction approach, it would require actuators and a large plate at the distal facet, adding undesired bulk and compromising the ultra-thin form factor. Here, we propose instead that switching be achieved through modulating wavelength using the concept of a reflector stack (Figure \ref{fig:reflectorDesign}a) with different reflectance behaviour at several closely spaced wavelengths.

The envisaged stack comprises layers of long-pass optical filters so that light penetrates to different depths at different wavelengths (Figure \ref{fig:reflectorDesign}b). Though the first-order reconstruction approach must be used on account of the wavelength modulation, it is still advantageous to satisfy the requirements for the zeroth-order model so that it can be used as a starting point for optimisation and because it is a special case of the first-order model.  Therefore, we require that the stack should couple light between different modes of the fibre, including polarisation modes, in a pseudo-random manner to ensure with high probability that the eigenvalues of the TM are distinct as required (Section \ref{subsec:zeroordermodel}). This could be realised by placing atop each filter a thin (20--50nm) layer comprising spatially heterogenous wire-grid polarisers (Figure \ref{fig:reflectorDesign}c) fabricated, for example, using elctron beam lithography \cite{Williams2016}.  To act as optical polarisers the fabricated wires must have width and pitch less than the wavelength of light ($<\lambda/2$) and so these layers are classed `metasurfaces'. Spatial heterogeneity would be achieved using arrays of partial polariser cells that have varying linear diattenuation (or dichroism) and diattenuation axis orientations (Figure \ref{fig:reflectorDesign}c).

Because of the optical filters, the reflector stack provides a different PM at each wavelength, $\mathbf{R}_n,~n=1.. 4$.  This is seen by considering the different sum of wavefronts that occurs due to reflection and absorption from different layers in Figure \ref{fig:reflectorDesign}a. Sweeping a tunable laser effectively switches between reflectors without any actuation at the distal facet.  The solid nature of these structures means they can be characterised before use, for example using the process in Appendix \ref{subsubsec:charBeforeUseAppendix}, and are expected to remain stable throughout operation. The linear nature of this reflector stack design ensures that each PM, $\mathbf{R}_n,~n=1.. 4$, is symmetric due to optical reciprocity \cite{Gu2015}.

\subsection{Selection of wavelengths}
\label{subsec:selectingWavelengths}
For a given reflector stack the optimal filter centre wavelengths, RM characterisation wavelengths and imaging wavelength must be determined.  We assume here that all filters are long-pass but a system using short-pass filters would also be feasible. First, we approximate the long-pass filter transmission function with a sigmoid:

\begin{equation}
\label{eq:sigmoidFilter}
f(\lambda, \lambda_{filt}) = \frac{\tau_{max}}{1 + e^{-\frac{\alpha}{\tau_{max}}\left(\lambda - \lambda_{filt}\right)}}
\end{equation}

\noindent where $\tau_{max}$ is the maximum power transmission (fixed here as 0.8 based on real component data), $\alpha$ determines the steepness of the filter (measured to be $\sim$0.73 based on numerous commercially available filters, e.g. ThorLabs FELH0850, which represents a $\sim$5nm wavelength change for 10\% to 90\% normalised transmission), and $\lambda_{filt}$ is the centre wavelength of the filter.  For the case of 3 RM characterisation wavelengths ($\lambda_1 < \lambda_2 < \lambda_3$), this leads to the following matrix:

\begin{equation}
\mathbf{F} = \left(\begin{array}{c c c}
f(\lambda_1, \lambda_a) & f(\lambda_1, \lambda_b) & f(\lambda_1, \lambda_c) \\
f(\lambda_2, \lambda_a) & f(\lambda_2, \lambda_b) & f(\lambda_2, \lambda_c) \\
f(\lambda_3, \lambda_a) & f(\lambda_3, \lambda_b) & f(\lambda_3, \lambda_c) \\
\end{array}\right)
\end{equation}

\noindent where $\lambda_a < \lambda_b < \lambda_c$ are the centre wavelengths of the 3 filters with transmission curves described by Equation \ref{eq:sigmoidFilter}.  To maximise the difference between the filters' behaviour at the interrogation wavelengths, we minimise the condition number of $\mathbf{F}$, which gives our objective function, $g$:

\begin{equation}
g(\lambda_1, \lambda_2, \lambda_3, \lambda_a, \lambda_b, \lambda_c) = \kappa(\mathbf{F}) = \lVert \mathbf{F}^{-1}\rVert_2 \lVert \mathbf{F}\rVert_2
\end{equation}

\noindent where $\kappa(..)$ represents condition number and $\lVert ..\rVert_2$ is the $\ell_2$ norm. We also want to ensure that at the imaging wavelength $\lambda_4$ ($>\lambda_3$) the reflector stack transmits significantly more light than at the longest characterisation wavelength ($\lambda_3$).  This is essential for efficient illumination and imaging.  To achieve this, we define a ratio, $\beta$:

\begin{equation}
\beta = \frac{f(\lambda_4, \lambda_c)}{f(\lambda_3, \lambda_c)}
\end{equation}

The full optimisation problem is then:

\begin{equation}
\begin{array}{c}
\min\limits_{\lambda_1, \lambda_2, \lambda_3, \lambda_a, \lambda_b, \lambda_c} g(\lambda_1, \lambda_2, \lambda_3, \lambda_a, \lambda_b, \lambda_c)\\
\mathrm{s.t.~~~~} \lambda_{min} < \lambda_1 < \lambda_2 < \lambda_3 < \lambda_4 < \lambda_{max} \\
\lambda_2 - \lambda_1, \lambda_3 - \lambda_2, \lambda_4 - \lambda_3 > \Delta \lambda_{las},~~~~\lambda_4 - \lambda_1 < \Delta\lambda_{fib}\\
\lambda_{min} < \lambda_a < \lambda_b < \lambda_c < \lambda_{max},~~~~\beta > \beta_{min}
\end{array}
\end{equation}

\noindent where $\lambda_{min}$ and $\lambda_{max}$ are the minimum and maximum range of the tunable laser, $\Delta \lambda_{las}$ is the minimum tuning step of the laser, and $\Delta \lambda_{fib}$ is the spectral bandwidth of the fibre.

Here, we set $\lambda_{min}$ and $\lambda_{max}$ to near-infrared wavelengths of 840nm and 860nm respectively \cite{Gordon2018}.  We set $\Delta \lambda_{las}$ to 0.2nm to reflect the tuning step of commercial tunable lasers.  $\Delta \lambda_{fib}$ is set to be 7nm representing a reasonable spectral bandwidth for a fibre of length $\sim$2m \cite{Carpenter2015, Thompson2012}. Finally, we choose $\beta_{min} = 10$ as providing acceptable isolation between characterisation and imaging.  With the objective and constraints defined, optimisation is performed using a genetic algorithm, implemented in MATLAB 2016b.

\subsection{Simulated fibre matrices}
\label{subsec:simulatingFibreMats}
For a first proof-of-principle, fibre TMs of size 32$\times$32 are simulated. This corresponds to 16 pixels of spatial resolution in two orthogonal polarisations.  The matrices are based on a model of a MCF with large core-to-core coupling. The 16 spatial pixels are considered to represent light guiding cores organised in a rectangular $4 \times 4$ grid with unit spacing (arbitrary units) between adjacent elements. The power coupling between a given core at the input and a given core at the output is modelled as decreasing exponentially with the squared lateral distance between them, i.e.:

\begin{equation}
p_{\alpha,\beta} = e^{\left(-\frac{\left(x_\alpha - x_\beta\right)^2 + \left(y_\alpha - y_\beta\right)^2}{\sigma_{fib}^2}\right)}
\end{equation}

\noindent where $p_{\alpha,\beta}$ is the power coupled between core $\alpha$ at the input (with coordinates $(x_\alpha, y_\alpha)$) and core $\beta$ at the output (with coordinates $(x_\beta, y_\beta)$).  $\sigma_{fib}^2$ is a coupling parameter which we set to be 3. This is a Gaussian power coupling model, which reflects empirical observations of real MCFs \cite{Gordon2019a}. The power coupling profile is duplicated for the second polarisation via a Kronecker product with a $2\times2$ matrix of ones. The complex phase of each element is drawn randomly from a uniform distribution, $\theta_{\alpha,\beta,p,s} \sim U(-\pi, \pi)$ where $p=1,2,~~s=1,2$ indicate the input and output polarisations respectively. This gives a combined expression for each matrix element:

\begin{equation}
a_{\alpha,\beta, p, s} = e^{\left(\frac{\left(x_\alpha - x_\beta\right)^2 + \left(y_\alpha - y_\beta\right)^2}{\sigma_{fib}^2}\right)} \cdot e^{i\theta_{\alpha,\beta,p,s}}
\end{equation}

The simulated TM, $\mathbf{A}$, is formed of all elements $a_{\alpha,\beta, p, s}$ in some ordering such that each pair ($\beta$, $s$) defines a unique row index, and each pair ($\alpha$, $p$) defines a unique column index.  $\mathbf{A}$ is then decomposed using singular value decomposition:

\begin{equation}
\mathbf{A} = \mathbf{U}_{A} \mathbf{S}_{A} \mathbf{V}_{A}^H
\end{equation}

\noindent where $\mathbf{U}_{A}$ and $\mathbf{V}_{A}$ are unitary, $\mathbf{S}_{A}$ is a diagonal matrix containing singular values of $\mathbf{A}$, and $(..)^H$ represents a Hermitian transpose. To ensure $\mathbf{A}$ is non-unitary, new singular values, $s_n,~n=1.. 32$, are generated such that $s_n = n/32,~n=1.. 32$. This ensures an invertible non-unitary matrix.

If using the zeroth-order model the simulated matrix is used as the TM at all wavelengths. If using the first-order model the simulated matrix is used as the TM at wavelength $\lambda_1$, $\mathbf{A}_1$, and the TMs at the other wavelengths are generated from the matrix logarithm of $\mathbf{A}_1$ ($d\mathbf{A}$ from Equation \ref{eq:dAfromA1}) and then applying Equations \ref{eq:firstOrderA2}--\ref{eq:firstOrderA4} to obtain $\mathbf{A}_2$, $\mathbf{A}_3$, $\mathbf{A}_4$.

Although this method of generating matrices simulates a MCF, it could just as easily represent a MMF through an appropriate change of mode basis, e.g. by redefining the spatial pixels to be Laguerre--Gauss modes \cite{Gataric2018, Carpenter2014}. Simulated TMs are generated using custom code written in MATLAB 2016b.

\subsection{Measured fibre matrices}
The proposed methodology is further validated by testing the first-order model and associated recovery using TMs measured from two real fibres.  The first is a MCF with TM of size $1648 \times 1648$ measured at a single wavelength of 850nm in \cite{Gordon2018}.  This TM is used as $\mathbf{A}_1$ and TMs at other wavelengths ($\mathbf{A}_2$, $\mathbf{A}_3$, $\mathbf{A}_4$) are generated from this by finding the matrix logarithm of $\mathbf{A}_1$ ($d\mathbf{A}$ from Equation \ref{eq:dAfromA1}) and then applying Equations \ref{eq:firstOrderA2} -- \ref{eq:firstOrderA4} to obtain $\mathbf{A}_2$, $\mathbf{A}_3$, $\mathbf{A}_4$. These are subsequently used to generate RMs from which the TMs are recovered. This process of adjusting TMs and generating RMs is implemented using custom code written in MATLAB 2016b. This case validates that the recovery method is feasible for large TMs. 

The second is a 2m piece of step-index MMF with $420 \times 420$ TMs measured across a wavelength range 1525 -- 1567nm in steps of 0.08nm (taken from \cite{Carpenter2016a}).  Because of the step-index nature of the fibre, the higher-order modes have relatively narrow spectral bandwidth (see Appendix \ref{sec:MMFBWcalc}).  Therefore, to improve bandwidth performance only the 110 lowest-order modes are used, giving submatrices of size $110 \times 110$.  Here, we use a wavelength spacing of $\Delta \lambda = 0.9$nm giving TMs measured at characterisation wavelengths $\lambda_1 = 1525.6$nm, $\lambda_2 = 1526.5$nm and $\lambda_4 = 1527.4$nm, and imaging wavelength $\lambda_4 = 1528.3$nm.  This wavelength separation is sufficient to enable the use of multiple off-the-shelf optical filters in the reflector stack. 

\subsection{Simulating reflectors}
\label{subsec:simulatingReflectors}
When recovering simulated TMs of size $2M \times 2M$ ($2M=32$ here, as described in Section \ref{subsec:selectingWavelengths}), we simulate reflectors by first creating a diagonal matrix with integers 1 to $2M$ along the main diagonal in some permutation.  Distinct permutations are used for the main diagonals of the three separate reflectors. Random permutations of the integers 1 to $2M-1$ are then used to create the subdiagonals and superdiagonals of these reflector matrices, mimicking the mode-coupling that would be expected in real reflectors. The singular value decomposition of this matrix is computed, and the matrix is then reconstructed from the left- and right-singular vector matrices but with the singular values replaced by a new set, $s_m = m/2M$ with $m=1..2M$. This ensures that the requirement for distinct eigenvalues  (see Section \ref{subsec:zeroordermodel}) is satisfied because of the square-root relationship between singular values and eigenvalues.

Next, we present a method of simulating physically realistic reflectors like that shown in Figure \ref{fig:reflectorDesign}a. This is achieved by extending the propagation matrix method used for simulating multilayer optical materials (Bragg reflectors or stacks) \cite{Orfanidis2016}. The reflector is modelled as a series of layers, as shown in Figure \ref{fig:reflectorDesign}a, and the propagation operator used is 2D Fresnel propagation, applied separately to each polarisation.  This is in contrast to conventional propagation matrix method in which the propagation operator is the complex exponential propagation operator $e^{-ikz}$ where $z$ is distance and $k$ is wavenumber. The Fresnel propagation operator enables computation of output fields $\mathbf{p}_{H}'$ and $\mathbf{p}_{V}'$ (where subscripts $H$ and $V$ denote horizontal and vertical polarisations respectively), resulting from input fields $\mathbf{p}_{H}$ and $\mathbf{p}_{V}$ propagating a distance $l$ through a layer as:

\begin{equation*}
\mathbf{p}_{H}' = \mathcal{F}^{-1} \left(\mathcal{F} \left(\mathbf{p}_{H}\right) e^{-i\frac{2\pi^2}{k} \left(\xi_x^2 +\xi_y^2\right) l}\right)
\end{equation*}

\noindent where $\mathcal{F}$ is the discrete Fourier transform, $n$ is the refractive index of the layer, $\lambda$ is the wavelength, $k = \frac{2\pi n}{\lambda}$, $\xi_x$ and $\xi_y$ are coordinates in the Fourier plane, and constant factors are neglected \cite{Goodman1996}. This is repeated for $\mathbf{p}_{V}$, giving a 2D complex vector at each point. 

Using this modified method, the reflector stack is simulated using three types of layer: pure glass, glass with a wire-grid polariser metasurface on the top, and optical filters. Here, we simulate absorptive optical filters but reflective filters would work equally well because their behaviour in the passband is the same. The filters have transmission curves typical of commercially available components and the centre wavelengths are determined using the process in Section \ref{subsec:selectingWavelengths}.

The stack simulated here comprises a first 1mm thick layer of glass, followed by the 3 filters in succession, each 3mm thick.  The metasurfaces, negligibly thin for propagation purposes, are placed between the glass and the first filter, between the first and second filters, and between the second and third filters.  The glass and filters have matched refractive indices of 1.52 and for simplicity we neglect the imaginary part of refractive index, i.e. distance-dependent loss.  Therefore, the only internal reflections occur at the metasurfaces, reflective filters (if used), and the final glass-air interface. Each metasurface is simulated by randomly generating diattenuating partial-polariser Jones matrices for each sampled point of the field.  Each point is assigned a diattenuation angle drawn from a uniform distribution, $\theta_D \sim U(-\pi,\pi)$, and a diattenuation drawn from a uniform distribution, $D \sim U(0,1)$.  The transmitted light at each pixel is computed by multiplying the sampled field vector by the relevant Jones matrix:

\begin{equation*}
\left(\begin{array}{c}
p_{tH} \\
p_{tV} \end{array}\right) = \sqrt{\alpha}
\left(\begin{array}{c c}
\cos \theta_D & -\sin \theta_D \\
\sin \theta_D & \cos \theta_D
\end{array}\right)
\left(\begin{array}{c c}
1 & 0 \\
0 & \sqrt{\frac{1-D}{1+D}}
\end{array}\right)
\end{equation*}
\begin{equation}
\cdot \left(\begin{array}{c c}
\cos \theta_D & \sin \theta_D \\
-\sin \theta_D & \cos \theta_D
\end{array}\right)
\left(\begin{array}{c}
p_{H}' \\
p_{V}' \end{array}\right)
\end{equation}

The light reflected is then given by:

\begin{equation*}
\left(\begin{array}{c}
p_{rH} \\
p_{rV} \end{array}\right) = \sqrt{\alpha}
\left(\begin{array}{c c}
\cos \theta_D & -\sin \theta_D \\
\sin \theta_D & \cos \theta_D
\end{array}\right)
\left(\begin{array}{c c}
0 & 0 \\
0 & 1 - \sqrt{\frac{1-D}{1+D}}
\end{array}\right)
\end{equation*}
\begin{equation}
\cdot \left(\begin{array}{c c}
\cos \theta_D & \sin \theta_D \\
-\sin \theta_D & \cos \theta_D
\end{array}\right)
\left(\begin{array}{c}
p_{H}' \\
p_{V}' \end{array}\right)
\end{equation}

\noindent where $\alpha$ is a parameter representing overall power loss in the metasurface due to plasmonic losses, set to a typical value of $~0.8$ here \cite{Williams2016}.  

The next step is to determine the overall PMs at the imaging and characterisation wavelengths, $\mathbf{R}_n,n=1...4$, and the TM of the reflector at the imaging wavelength, $\mathbf{A}_{refl}$.  This is achieved by iteratively propagating different input fields forwards and backwards through the stack until a steady-state solution is reached using custom code written in MATLAB.  This process is repeated at each of the desired wavelengths. The coupling between spatial modes introduced by diffractive propagation between layers and the coupling between polarisations introduced by the metasurfaces produces PMs that with high probability have distinct eigenvalues, which is necessary for approximate zeroth-order TM recovery that precedes first-order TM recovery (Section \ref{subsec:zeroordermodel}).

The input fields chosen may be derived from the columns of the fibre TM (the modes of the fibre), because any field exiting the fibre must be a linear combination thereof.  However, the modes of the fibre may be expressed in some particular basis, e.g. Laguerre--Gauss modes for MMF, and so must first be converted to Cartesian basis so that Fresnel propagation can be applied. The light coupled back into the fibre can only be coupled into a linear combination of the reverse-propagating (i.e. complex conjugated) fibre modes. Therefore, the steady-state reflected field is re-expressed in the fibre mode basis.  This means the reflector matrices have the same size as the fibre TM ($\in \mathbb{C}^{2M \times 2M}$), but $\mathbf{A}_{refl}$ will in general be $\in \mathbb{C}^{2N \times 2M}$. Because the $2N$-dimensional vector output of $\mathbf{A}_{refl}$ is in a Cartesian pixel basis, it is straightforward to compute sample illumination as per Section \ref{subsec:imaging}.

Both methods of simulating reflector PMs are implemented in custom written code in MATLAB 2016b.

\subsection{Noise model}
\label{subsec:noiseModel}
We now consider the effect of noise on the presented TM characterisation and imaging system. Noise is inherently present in any realistic implementation of this system, such as that proposed in Appendix \ref{subsec:expSetupAppendix}, and so must be examined to demonstrate feasibility and robustness.

The main source of noise in such measurements (for example data acquired using the system in Figure \ref{fig:proposedExpSetup}) is assumed to be electronic noise introduced by the image sensor.  However, there are many processing steps (mostly matrix operations) that transform these raw images into the RMs used for reconstruction \cite{Gordon2018}. The noise in each image pixel is assumed to be independent, which is consistent with previous fibre imaging work \cite{Gordon2018,Gordon2019a}. By the central limit theorem,  the repeated multiplication and addition of noisy variables via matrix operations will produce circularly symmetric complex zero-mean Gaussian noise.   The relatively strong optical confinement (i.e. wave-guiding) during RM characterisation will result in large detected optical power and therefore approximately Gaussian shot noise. Laser power fluctuation (relative intensity noise) is not inherently Gaussian, but after time averaging and repeated matrix operations creates a Gaussian contribution to overall noise \cite{Yamamoto1983}.

Therefore, we model the noise in the system by adding random matrices with independent circularly symmetric complex Gaussian elements to each PM ($\mathbf{R}_1$, $\mathbf{R}_2$, $\mathbf{R}_3$) and each RM ($\mathbf{C}_1$, $\mathbf{C}_2$, $\mathbf{C}_3$). For example, a noisy version of $\mathbf{R}_3$ is given by:

\begin{equation}
\mathbf{\hat{R}}_3 = \mathbf{R}_3 + \mathbf{N}_3
\end{equation}

\noindent where each element of $\mathbf{N}_3$ is a complex random variable, $n=a+ib$, drawn from the distributions:

\begin{equation}
\label{eq:noiseDistributions}
a \sim \mathcal{N}(0,\sigma_n)~~~~~~~~~~~~~~~b \sim \mathcal{N}(0,\sigma_n)
\end{equation}

The same distributions, parameterised by $\sigma_n$, are used to generate noise matrices $\mathbf{\hat{R}}_2$, $\mathbf{\hat{R}}_3$, $\mathbf{\hat{C}}_1$, $\mathbf{\hat{C}}_2$ and $\mathbf{\hat{C}}_3$. We vary $\sigma_n$ over a wide range, including values reflecting empirically observed experimental noise (see Figure \ref{fig:errorGraphWithNoise}) \cite{Gordon2018}.

To model noise in image reconstruction we add circularly symmetric complex Gaussian noise to $\mathbf{A}_{refl}$, $\mathbf{R}_4$ and the measured field at the proximal fibre facet, $\mathbf{y}_{total}$, to get noisy version $\mathbf{\hat{A}}_{refl}$, $\mathbf{\hat{R}}_4$ and $\mathbf{\hat{y}}_{total}$ respectively.  Combined with the noisy estimate of the TM at the imaging wavelength, $\mathbf{\hat{A}}_4$, we can then get a noisy estimate of the illuminated target, $\mathbf{\hat{r}}_{target}$.  A noisy estimate of the illumination can then be generated using $\mathbf{\hat{A}}_4$ and $\mathbf{\hat{A}}_{refl}$ using Equation \ref{eq:xTarget}, and we can recover a noisy estimate of $\mathbf{\hat{r}}_{target}$ using Equation \ref{eq:illumCorrectionProb}.  Simulation of noise is implemented using custom code in MATLAB 2016b.

\subsection{Performance metrics}
\label{subsec:performanceMetrics}
To validate the performance of the presented method, we define several error metrics. The first is the error between the recovered TM at the imaging wavelength, $\mathbf{\tilde{A}}_{4}$, and the ideal TM, $\mathbf{A}_4$, given as:

\begin{equation}
\label{eq:matError}
\epsilon_{mat} = \frac{||\mathbf{\tilde{A}}_{4} - \mathbf{A}_{4}||_F}{||\mathbf{A}_{4}||_F} = \frac{\left \lVert \left(\mathbf{\tilde{A}}_{4}^{-1} \mathbf{A}_{4} - \mathbf{I}\right) \mathbf{A}_{4} \right \rVert_F}{||\mathbf{A}_{4} ||_F}
\end{equation}

\noindent where $|| .. ||_F$ represents the Frobenius norm.  Dividing by the norm of the ideal matrix, $\mathbf{A}_4$ ensures the metric presents normalised (or proportional) error. We then define a second error metric that represents the error introduced when using a recovered matrix to reconstruct an image:

\begin{equation}
\label{eq:imError}
\epsilon_{image} = \frac{||\mathbf{\tilde{A}}_{4}^{-1} \mathbf{A}_{4} \mathbf{r}_{target} - \mathbf{r}_{target}||_2}{||\mathbf{r}_{target} ||_2} 
=
\frac{\left \lVert \left(\mathbf{\tilde{A}}_{4}^{-1} \mathbf{A}_{4} - \mathbf{I}\right) \mathbf{r}_{target} \right \rVert_2}{||\mathbf{r}_{target} ||_2}
\end{equation}

\noindent where $\mathbf{r}_{target}$ is the ideal image. A further metric measures error due to noise in image recovery:

\begin{equation}
\label{eq:imError_n}
\epsilon_{image,n} = \frac{||\mathbf{\tilde{r}}_{target} - \mathbf{r}_{target}||_2}{||\mathbf{r}_{target} ||_2}
\end{equation}

\noindent where $\mathbf{\tilde{r}}_{target} $ is the estimated recovered image as defined in Equation \ref{eq:imRecovery}.

\subsection{Recovering TMs}
Having produced the necessary instantaneous RMs ($\mathbf{C}_n,~n=1..3$), based on real or simulated TMs, the next step is to solve for the TMs using \emph{a priori} known reflector PMs, $\mathbf{R}_n$ with $n=1..3$.  For TMs and RMs generated using the zeroth-order model, the equations of Section \ref{subsec:zeroordermodel} are solved using code written in MATLAB 2016b, including solving the systems of linear equations generated by Equation \ref{eq:schurSolution}.

In the first-order case, the recovery algorithm depicted in Figure \ref{fig:firstOrderFlowchart} is implemented in the TensorFlow package via Python because of its efficient GPU utilisation and hence superior speed.  Initially, we use the Adam stochastic gradient descent optimiser \cite{Kingma2014}, changing to a conventional gradient descent optimiser for fine adjustment in the last few iterations.

The zeroth-order recovery algorithm typically finishes in less than 1s.  However, first-order iterative algorithm is significantly slower and can take several hours for 1500 iterations on a $1648 \times 1648$ matrix on a single Tesla K40 GPU. For smaller $32 \times 32$ matrices requiring about 500 iterations, the running time is typically $<5$ seconds. 

\section{Results}
\label{sec:results}
\subsection{Simulated TMs}
\begin{figure}[htpb!]
	\centering
	\includegraphics[height=0.7\textheight]{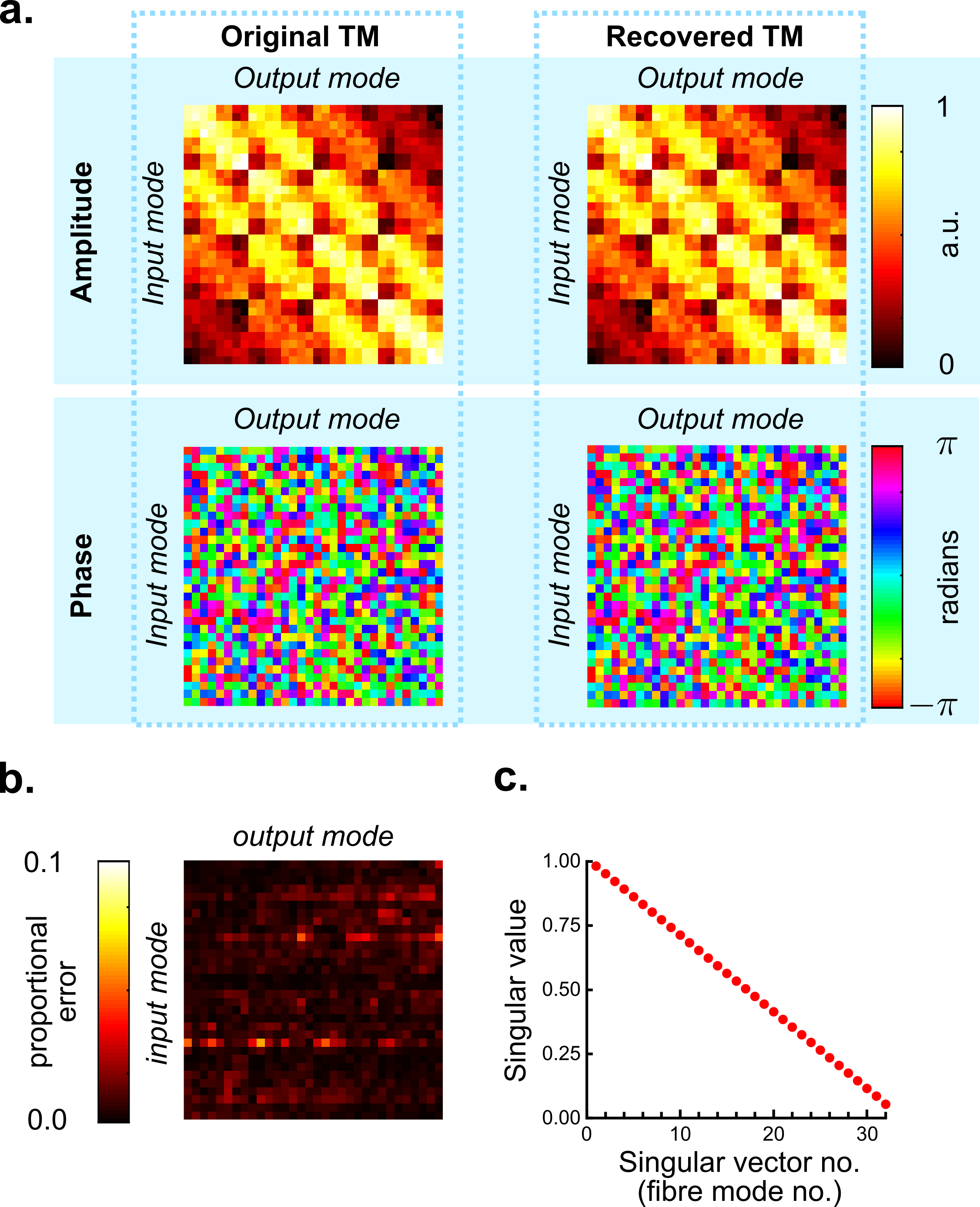}
	\caption{\textbf{Recovery of a simulated TM using the first-order recovery algorithm}: a) The original simulated TM is compared with the recovered TM (noise level $\sigma_n = 10^{-3}$) demonstrating successful recovery. b) Proportional element-wise error in TM reconstruction (maximum $<0.05$) after 300 iterations of first-order model. c) Singular values of TM demonstrating non-unitarity.}  
	\label{fig:exampleSimulatedMatrix}
\end{figure}

We first randomly generate a single $32 \times 32$ non-unitary matrix, $\mathbf{A}_1$, and then $\mathbf{A}_2$, $\mathbf{A}_3$, $\mathbf{A}_4$ using the first-order model. We then randomly generate reflector PMs $\mathbf{R}_n$ with $n=1...4$ and use $\mathbf{A}_1$ to simulate RMs $\mathbf{C}_n$ with $n=1...3$ and perform TM recovery using the first-order method (Figure \ref{fig:exampleSimulatedMatrix}a) with element-wise error $<0.05$ (Figure \ref{fig:exampleSimulatedMatrix}b). The singular values of this matrix vary from 1 down to 0.03 (Figure \ref{fig:exampleSimulatedMatrix}c), verifying that this recovery can be performed on very non-unitary matrices.  

\begin{figure}[htpb!]
	\centering
	\includegraphics[height = 0.65\textheight]{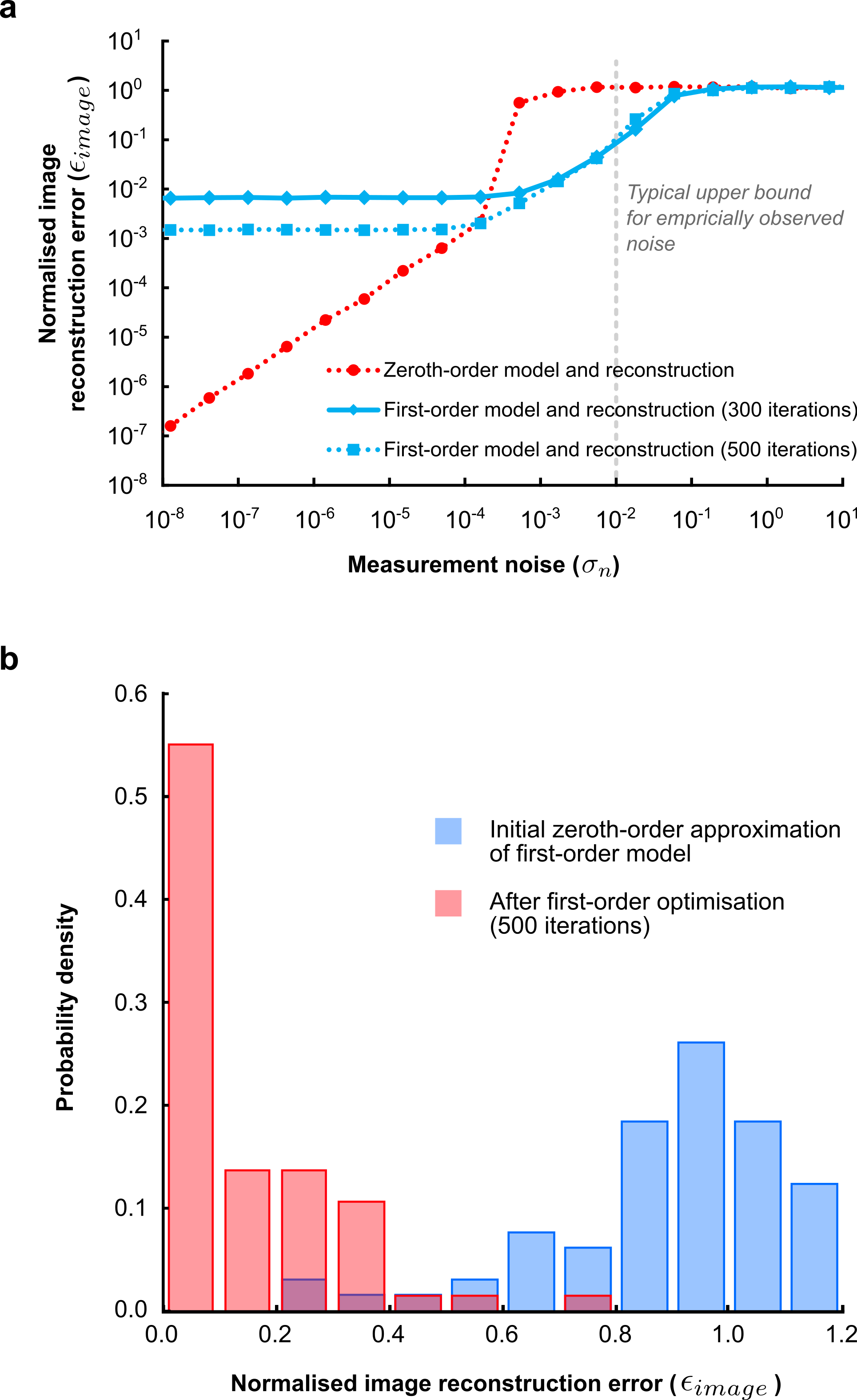}
	\caption{\textbf{Impact of measurement noise and different perturbation conditions on recovery}. a) Error in reconstructed images for different levels of noise in measured quantities ($\epsilon_{image}$ and $\sigma_N$ are defined in Section \ref{subsec:performanceMetrics}).  At low noise levels the error of the first-order model is seen to be limited by the number of iterations, but at higher noise levels converges to a trend consistent with the zeroth-order model.  The first-order approach is more robust at higher noise levels due to the use of iterative optimisation. A typical upper bound for noise encountered experimentally \cite{Gordon2018} is indicated to show that recovery from real experimental data is feasible. b) Histogram showing effect of 200 different random TM realisations, i.e. different fibre perturbations, under the first-order model.  Error after the initial zeroth-order approximation (see Figure \ref{fig:firstOrderFlowchart}) can be large ($>1$) but after 500 iterations, 50\% of TMs have errors $<0.1$ and 95\% have errors $<0.4$.}
	\label{fig:errorGraphWithNoise}
\end{figure}

Next, we examine the impact of noise by repeating this recovery process for 64 different noise realisations with increasing noise power (see Section \ref{subsec:noiseModel}). Matrices are generated using either the zeroth- or first-order model and then reconstructed using the respective recovery algorithm.  The resultant error in image reconstruction using the recovered TM (Figure \ref{fig:errorGraphWithNoise}a) for the zeroth-order model, in which the recovery is largely analytical, deteriorates rapidly in performance beyond a noise threshold (visualised as a `hump' in Figure \ref{fig:errorGraphWithNoise}a). In the first-order model, in which recovery relies on an optimisation process, such instability is not observed, thus it is suspected to arise from numerical error accumulation when computing analytical results. The first-order model exhibits improved stability at higher noise levels (such as those encountered experimentally \cite{Gordon2018}) because the optimisation process, by definition, does not terminate until error (including numerical error) is minimised. At lower noise levels, first-order performance is limited by the number of iterations of the algorithm, which is evident from the impact of increasing the number of iterations from 300 to 500.  

Furthermore, testing the effect of 200 different random TM realisations that could result from perturbations of the fibre or manufacturing variations (Figure \ref{fig:errorGraphWithNoise}b) shows that after 500 iterations using the first-order recovery algorithm, the error converges towards zero with $>$50\% of matrices having error $< 0.1$ and 90\% having error $<0.4$. Negligibly small noise power is used so as to only observe the effect of perturbations. 

\subsection{Real MCF TMs}
\label{subsec:realMCFresults}
Having established the potential of the first-order model using simulated matrices, we then applied it to an experimentally measured MCF TM \cite{Gordon2018}. Using the optimisation procedure of \ref{subsec:selectingWavelengths}, we determine the wavelengths to be used as follows: filters centred at 840.0nm, 842.7nm and 849.3nm, characterisation wavelengths of $\lambda_1$ = 843.4nm, $\lambda_2$ = 844.9nm and $\lambda_3$ = 846.5nm, and imaging wavelength $\lambda_4$ = 850.4nm.  

\begin{figure}[htpb]
	\centering
	\includegraphics[height=0.8\textheight]{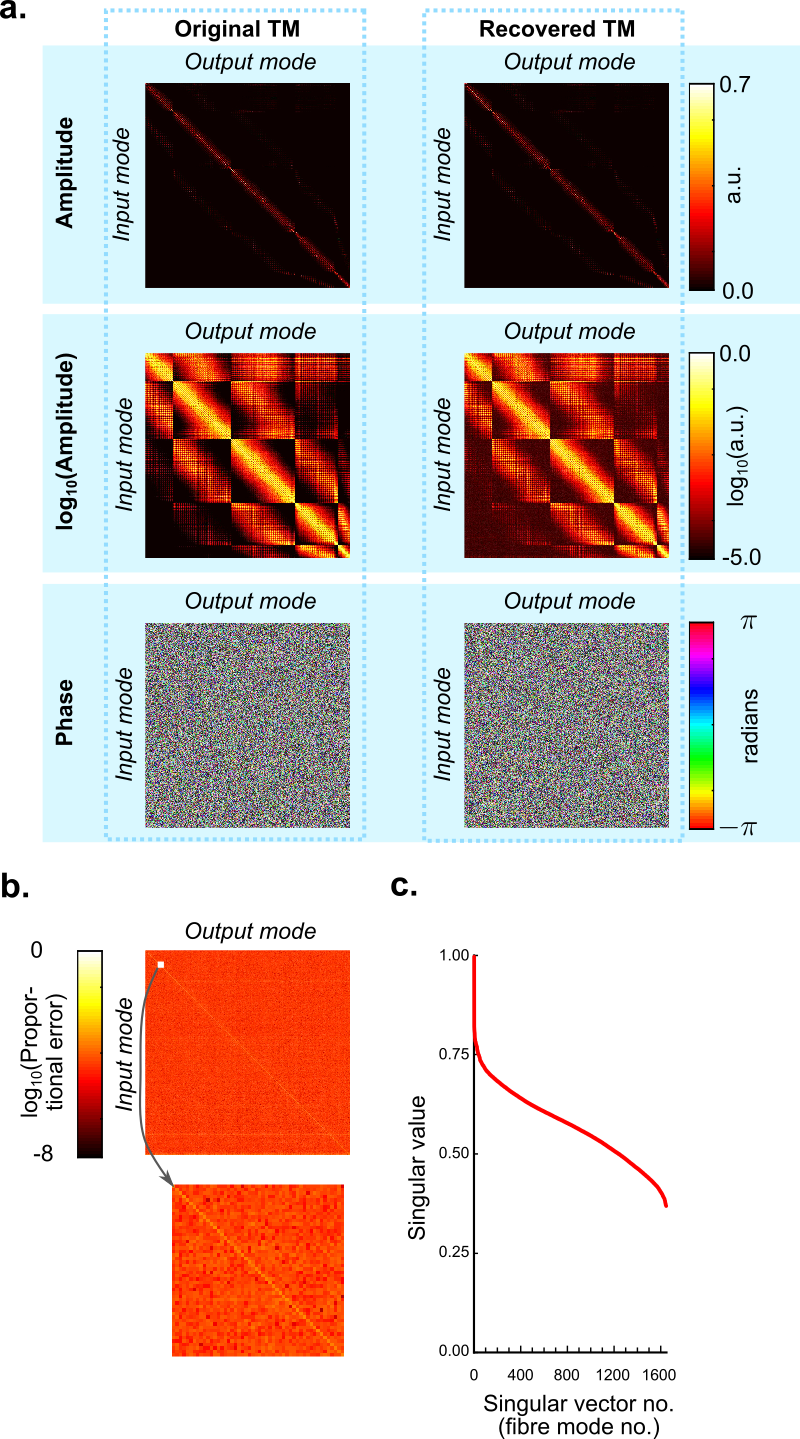}
	\caption{\textbf{First-order recovery of an experimentally-measured MCF TM}. a) Original TM using data from \cite{Gordon2018} and the first-order recovered TM, showing high visual similarity indicating successful recovery. b) Proportional element-wise error in TM reconstruction (maximum $<10^{-3}$). c) Singular values of TM demonstrating non-unitarity.}
	\label{fig:exampleRealMatrix}
\end{figure}

The non-square TM is first `downsampled' to a square form of size $1648 \times 1648$ (Figure \ref{fig:exampleRealMatrix}a). This downsampling process can be achieved via many methods, for example that detailed in Appendix \ref{subsec:downsamplingAppendix}, and can be reversed at the end of recovery to obtain an estimate for the original non-square TM.

The measured TM is used as $\mathbf{A}_1$ and we apply the first-order model to simulate $\mathbf{A}_2$, $\mathbf{A}_3$ and $\mathbf{A}_4$. Next, realistic reflector PMs, $\mathbf{R}_n$ with $n=1...4$, are simulated (see Section \ref{subsec:simulatingReflectors}) and RMs $\mathbf{C}_1$, $\mathbf{C}_2$ and $\mathbf{C}_3$ computed using Equations \ref{eq:C1}--\ref{eq:C3}.  Using the first-order algorithm we then recover the TM at the imaging wavelength, $\mathbf{A}_4$ (Figure \ref{fig:exampleRealMatrix}a). After 1500 iterations, the maximum element-wise error of the recovered matrix is $<10^{-3}$ (Figure \ref{fig:exampleRealMatrix}b). Because of the dense core spacing ($<4\mu$m) there is significant coupling between cores at the wavelength used, 852nm \cite{Gordon2018}, meaning the MCF begins to exhibit mode-coupling properties more like MMF. The singular values (Figure \ref{fig:exampleRealMatrix}d) show that the matrix is non-unitary but has a low condition number ($\sim$4), again similar to experimental measurements of MMF \cite{Carpenter2014}. 

\begin{figure}[htpb]
	\centering
	\includegraphics[height=0.7\textheight]{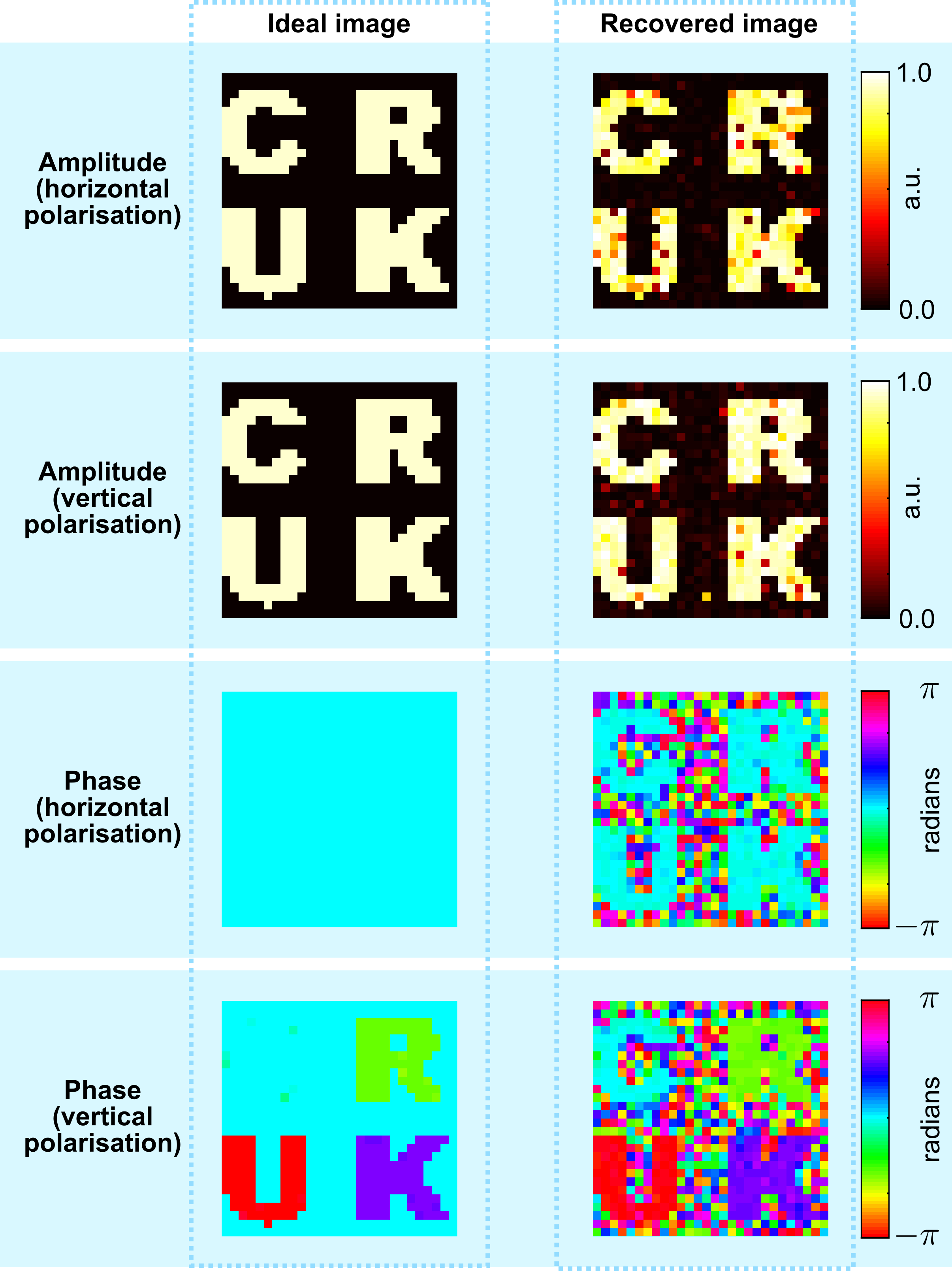}
	\caption{\textbf{Successful reconstruction of an amplitude, phase and polarisation image of a target at the distal facet using a TM recovered using the first-order model.} The recovered image noise is largely driven by illumination correction (see Section \ref{subsec:imaging}).}
	\label{fig:exampleRealImaging}
\end{figure}

Having recovered the TM, we next test the imaging procedure using random illumination. A $28 \times 28$ image is simulated and then successfully recovered (Figure \ref{fig:exampleRealImaging}). The image, $\mathbf{r}_{target}$ with reference to Figure \ref{fig:physicalModel}, comprises 28 $x$ positions $\times$ 28 $y$ positions $\times$ 2 polarisations $=$ 1568 degrees of freedom (i.e. $2N = 1568$ as defined in Section \ref{subsec:imaging}), which ensures that full reconstruction through a TM of dimension 1648$\times$1648 (i.e. $2M = 1648$ as defined in Section \ref{subsec:imaging}) is possible. The total image reconstruction error, $\epsilon_{n,image}$ (Equation \ref{eq:imError_n}), is $<0.1$ and arises primarily from illumination correction. Lower illumination levels result in higher noise (see Section \ref{subsec:imaging}) and could thus be improved by averaging over multiple random illumination conditions i.e. speckle averaging \cite{Choi2012}. Under conditions of perfectly uniform illumination, we calculate that $\epsilon_{n,image}$ becomes $<0.002$, giving a lower bound on error dominated by other facts e.g. computational limitations.

\subsection{Real multi-wavelength MMF TMs}
\label{subsec:realMMFtms}
\begin{figure}[htp!]
	\centering
	\includegraphics[height=0.7\textheight]{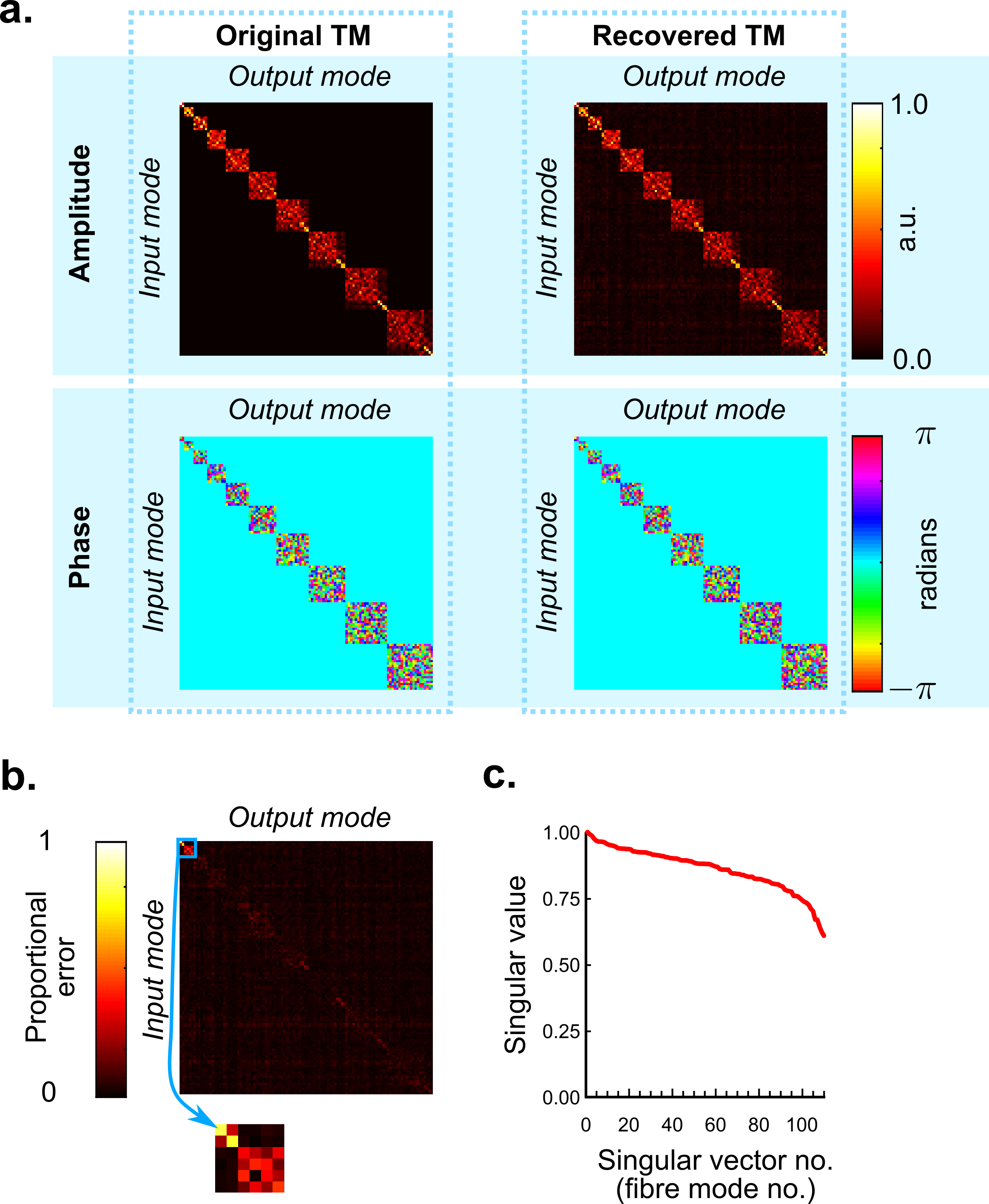}
	\caption{\textbf{First-order recovery of an experimentally-measured step-index MMF TM, using TMs recorded at multiple wavelengths (1525.6nm, 1526.5nm, 1527.4nm and 1528.3nm) \cite{Carpenter2016a}.} a) Original TM using data from \cite{Carpenter2016a} and the first-order recovered TM, showing high visual similarity indicating successful recovery. b) Proportional element-wise error in TM reconstruction values typically $<$0.05 except for the very lowest order modes that exhibit error $\sim$0.2 due to phase error. c) Singular values of measured TM demonstrating non-unitarity.}
	\label{fig:resultJoelMat}
\end{figure}

Using a multi-wavelength fibre TM measured from a 2m piece of step-index MMF \cite{Carpenter2016a} to generate RMs, $\mathbf{C}_n,~n=1..3$, also led to successful TM, $\mathbf{A}_4$, recovery (Figure \ref{fig:resultJoelMat}a). In this case, the characterisation wavelengths chosen are $\lambda_1 = 1525.6$nm, $\lambda_2 = 1526.5$nm and $\lambda_3 = 1527.4$nm, and the imaging wavelength is $\lambda_4 = 1528.3$nm. A $110 \times 110$ subset of the full TM data set is used in order to limit spectral bandwidth appropriately (see Appendix \ref{sec:MMFBWcalc}). After 475 iterations taking 14.5 minutes, the algorithm converges with significant visual similarity (Figure \ref{fig:resultJoelMat}a).  The matrix error, $\epsilon_{mat}$ (Equation \ref{eq:matError}) reaches a value of 0.26.  Looking at the element-wise error (Figure \ref{fig:resultJoelMat}b) we see that error is typically $<0.05$ except in the 3 lowest-order modes, in which the value reaches 0.2.  This error is observed to manifest as phase errors, with the amplitude values being largely accurate. Importantly, the singular value distribution agrees with that expected of MMF (Figure \ref{fig:resultJoelMat}c).

\section{Discussion}
The ability to form images through hair-thin optical fibres is currently limited due to the need for \emph{in situ} TM calibration. To overcome this limitation, we propose a method for recovering the instantaneous TM of a fibre based only on instantaneous reflection-mode measurements and \emph{a priori} characterisation reflection- and transmission-mode measurements. Calibration is achieved using three reflectors placed at the distal tip of the fibre, each of which produces a different reflection matrix based on a small modulation in wavelength, from which the imaging TM can be determined. As the fibre TM will differ slightly at each of these wavelengths, we developed a first-order method of modelling and compensating this change using matrix exponentials, which is valid within the spectral bandwidth of the fibre.  Under these assumptions, we demonstrate that it is possible to recover TMs and perform imaging using realistic optical components with acceptable noise tolerance for both simulated and measured TMs, including an experimentally measured MCF TM and a multi-wavelength MMF TM.

Nonetheless, several challenges must be overcome for experimental deployment of the proposed method. Given the demonstrated validity of the current model within the spectral bandwidth of the fibre, at 850nm the wavelength modulation would need to be within $\sim 5$nm for a typical 1m length of MCF   \cite{Thompson2012} or at 1550nm within $\sim 10$nm for a typical 2m length of MMF \cite{Carpenter2015}. In biomedical endoscopy, fibre lengths are typically 1--2m, however, for applications in industrial inspection, greater lengths may be needed. To extend the model to work for smaller bandwidth fibres, e.g. longer fibres or MMFs with very large numbers of modes, either the characterisation bandwidth could be reduced or a more complex (`high order') fibre model used.  The characterisation bandwidth is typically limited by the `sharpness' of available optical filters, which enables significant modulations in reflectance/absorbance over very small wavelength ranges.  Custom-fabricated filters may offer improved performance over off-the-shelf products.  Alternatively, more complex models of the fibre could be used.  Complete propagation models of graded-index MMF have been used to accurately compensate bending \cite{Ploschner2015} but require precise \emph{a priori} knowledge of the refractive index profile.  A major advantage of the approach presented here is that no such prior knowledge of the fibre is required, enabling the use of more complex refractive index profiles such as MCF.  It may be possible to develop a more complex model that uses the differential changes in the TM with respect to wavelength to model the fibre over a larger bandwidth for example using machine learning techniques \cite{Rahmani2018, Yang2018}.  As they do not require a reference model, machine learning techniques may also enable the use of a smaller reflector stack, with only 2 reflectors, though our preliminary investigations have not conclusively shown that this produces a unique solution when using an optimisation process to recover the TM.

Another challenge compared to existing transmission-mode fibre characterisation systems is the introduction of additional characterisation steps, both before and during imaging for instantaneous TM recovery (see Appendix \ref{subsec:operatingProcessApp}). For example, to recover a single TM will require 3 times as many measurements because of the 3 distinct wavelengths. Speed can be improved by experimental design, for example using high-speed digital micromirror devices instead of liquid-crystal SLMs \cite{Mitchell2016}, or by exploiting \emph{a priori} known sparsity structures of fibres to parallelise spatial measurements \cite{Gordon2018}.  Further, because the characterisation measurements use laser light at distinct wavelengths, they could be captured simultaneously by placing appropriate filters in front of 3 separate image sensors, one for each wavelength.  Computational recovery of TMs and images also presents a speed limitation that could be addressed by using additional GPUs or specialised hardware like an FPGA or ASIC, fine-tuning the algorithm and solver settings, or only reconstructing a single polarisation (a 4-fold speed increase).  If an analytical solution for the first-order case could be found, a dramatic speed increase would result.

A final challenge is fabricating and installing an appropriate reflector stack at the distal tip of the fibre. While optical filters can be purchased off-the shelf, wire-grid polariser metasurfaces most likely require custom fabrication using nanofabrication techniques such as electron-beam lithography.  Single-layer devices of this nature have already been demonstrated \cite{Williams2016, Williams2017}, but extending these to make reflector stacks will involve a number of additional deposition, lithography and processing steps. Though the metasurfaces must have specific polarisation transmission and reflection properties, they need not be fabricated to a specific design with high precision because they can be characterised \emph{in situ}.

Because the method does not place any requirements on the singular value distribution of the TM other than that the TM is invertible (i.e. unitarity is not required), it may also be applicable to more general scattering media. We provide further validation of this by demonstrating recovery of highly non-unitary simulated TMs with a very uneven distribution of singular values, similar to the quarter-circle distribution found in random scattering media \cite{Popoff2011a}. Nonetheless, it would require a precisely characterised reflector to be placed inside the medium, which would require an invasive step in tissue imaging applications.

\section{Conclusion}
In summary, we have developed a new approach to determine the transmission matrix of a multi-mode optical fibre without requiring access to the distal facet. We have demonstrated successful transmission matrix recovery and imaging using realistic optical components with acceptable noise tolerance for simulated transmission matrices, and those experimentally measured from a multi-core fibre and a multi-wavelength multi-mode fibre. The proposed method paves the way for experimental realisation of lensless imaging through hair-thin optical fibres.

\begin{acknowledgments}
GSDG acknowledges funding from Cancer Research UK (C47594/A21102, C55962/A24669); and a pump-priming award from the Cancer Research UK Cambridge Centre Early Detection Programme (A20976). CW acknowledges funding from a Cancer Research UK Pioneer Award (C55962/A24669). MG acknowledges funding from Engineering and Physical Sciences Research Council for the Centre for Mathematical and Statistical Analysis of Multimodal Clinical Imaging (EP/N014588/1).  RPM acknowledges funding from the Engineering and Physical Sciences Research Council (EP/L015889/1). SEB acknowledges funding from Cancer Research UK (C47594/A16267, C14303/A17197, C47594/A21102); and the EU FP7 agreement (FP7-PEOPLE-2013-CIG-630729).  GSDG would like to thank Peter Christopher from the University of Cambridge for useful discussions.
\end{acknowledgments}

\appendix
\section{Conceptual experimental implementation}
\label{sec:conceptualExpAppendix}
\subsection{Experimental set-up}
\label{subsec:expSetupAppendix}
\begin{figure}[htb!]
	\centering
	\includegraphics[width=\linewidth]{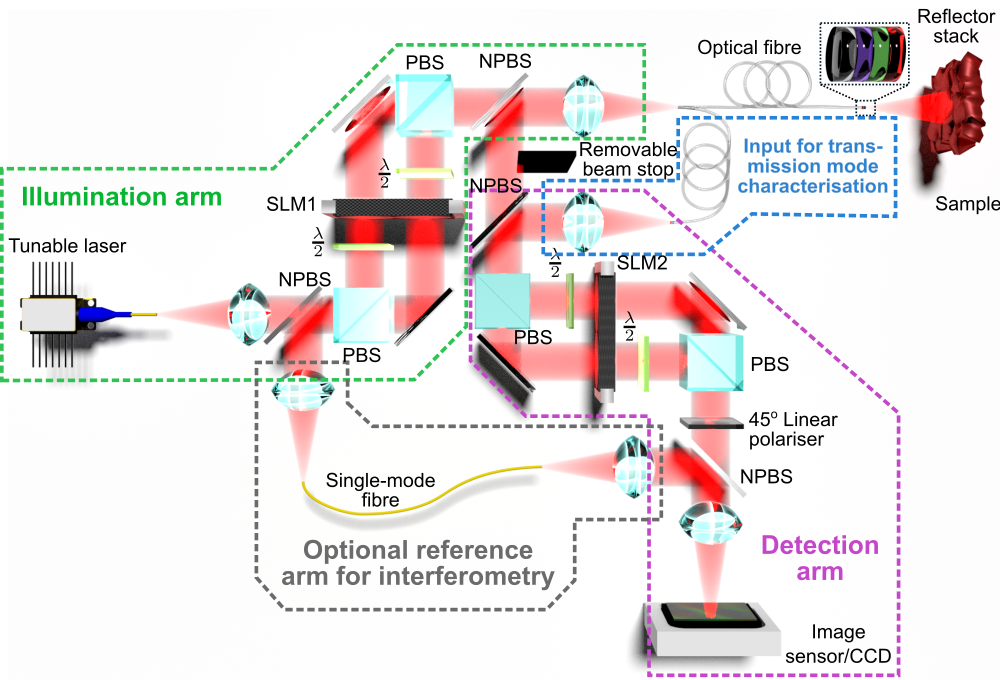}
	\caption{\textbf{Conceptual experimental set-up for fibre TM characterisation in reflection mode configuration using distal reflector stack.} SLM: spatial light modulator, PBS: polarising beam splitter, NPBS: non-polarising beam splitter, $\lambda/2$: half waveplate.}
	\label{fig:proposedExpSetup}
\end{figure}

A conceptual experimental set-up for reflection-mode TM recovery is shown in Figure \ref{fig:proposedExpSetup}.  There are two key sections: the illumination arm and the detection arm.  These perform similar functions to previous work \cite{Gordon2018} but are both located at the proximal fibre facet.  There is also an additional input that allows characterisation in transmission mode prior to use. Though our previous fibre TM characterisation set-ups have used non-interferometric phase retrieval \cite{Gordon2018}, an optional reference arm for interferometric imaging is included as it is widely used method in fibre characterisation \cite{Cizmar2012}.  The tunable laser is required to provide a high-coherence source at the different characterisation and imaging wavelengths.  These can typically be tuned in steps of 0.1--0.2nm, sufficient to cover a range of fibre TMs with different spectral bandwidths.

Proposed designs of a reflector stack are shown in Figure \ref{fig:reflectorDesign}. The plasmonic metasurface reflectors can be fabricated on a glass substrate using electron beam lithography patterning followed by metal deposition to create high-resolution wire-grid polarisers \cite{Williams2016}.  Fabrication could be scaled-up for commercial applications using, for example, UV lithography or interference lithography \cite{Neisser2015}. The absorptive or reflective long-pass filters used for the stack can be purchased off-the-shelf or made to order with sufficiently small gradations in centre wavelength. Stability of the stack over time is important to ensure the reflector matrices remain relatively constant. This can be achieved by using stable metals such as gold and depositing protective layers, e.g. MgF, on top of metasurfaces.

\subsection{Downsampling Reflection Matrices}
\label{subsec:downsamplingAppendix}
In practical implementations we may measure RMs that are non-square, denoted $\tilde{\mathbf{C}}_n$ with $n=1...4$, due to different sampling schemes for $\mathbf{x}$ and $\mathbf{y}$ of Figure \ref{fig:physicalModel}.  For example, the number of pixels on the camera may greatly exceed the number of different input calibration fields used to characterise the RM -- in our previous work characterising TMs there were $\sim 10^3$ times as many pixels as calibration inputs \cite{Gordon2018}.

In this scenario, consider $2M$ calibration samples $(\tilde{\mathbf{x}}_s, \tilde{\mathbf{y}}_s)_{s=1}^{2M}$ where $\tilde{\mathbf{x}}_s\in\mathbb{C}^{P\times 1}$ and $\tilde{\mathbf{y}}_s\in\mathbb{C}^{Q\times 1}$. Denoting $\tilde{\mathbf{C}}_n \in\mathbb{C}^{Q\times P}$ as a mapping of the fibre $\tilde{\mathbf{y}}=\tilde{\mathbf{C}}_n \tilde{\mathbf{x}}$ (with reference to Figure \ref{fig:physicalModel}), we wish to determine a downsampling process to obtain a square RM, $\mathbf{C}\in\mathbb{C}^{2M\times 2M}$ that may be used in our recovery algorithms.  One way of doing this is to estimate the largest singular values and corresponding left singular vectors of $\tilde{\mathbf{C}}_n \in\mathbb{C}^{Q\times P}$, which should match those of $\mathbf{C}_n \in\mathbb{C}^{2M\times 2M}$. This is done in the following way. Let

\begin{equation}
\tilde{\mathbf{X}}_\mathrm{cal}:=[\tilde{\mathbf{x}}_1, \ldots, \tilde{\mathbf{x}}_{2M}]\in\mathbb{C}^{P\times 2M}, 
\quad 
\tilde{\mathbf{Y}}_\mathrm{cal}:=[\tilde{\mathbf{y}}_1, \ldots, \tilde{\mathbf{y}}_{2M}]\in\mathbb{C}^{Q\times 2M}
\end{equation}

\noindent so that $\tilde{\mathbf{Y}}_\mathrm{cal}=\tilde{\mathbf{C}}\tilde{\mathbf{X}}_\mathrm{cal}$. If the coefficients of an input signal $\tilde{\mathbf{x}}\in\mathbb{C}^{P\times1}$ with respect to the $2M$-dimensional calibration basis are denoted by $\tilde{\mathbf{x}}_\mathrm{cal}\in\mathbb{C}^{2M\times1}$, i.e., $\tilde{\mathbf{x}}:=\tilde{\mathbf{X}}_\mathrm{cal}\tilde{\mathbf{x}}_\mathrm{cal}$, then its output signal is given as $\tilde{\mathbf{y}}=\tilde{\mathbf{C}}\tilde{\mathbf{x}}=\tilde{\mathbf{C}}\tilde{\mathbf{X}}_\mathrm{cal}\tilde{\mathbf{x}}_\mathrm{cal}=\tilde{\mathbf{Y}}_\mathrm{cal}\tilde{\mathbf{x}}_\mathrm{cal}$.
In other words, $\tilde{\mathbf{Y}}_\mathrm{cal}$ is a mapping of the fibre from the $2M$-dimensional calibration basis to $\mathbb{C}^{Q\times 1}$, and it is hoped that the largest $2M$ singular values and corresponding left singular vectors of  $\tilde{\mathbf{Y}}_\mathrm{cal}$ approximate those of $\tilde{\mathbf{C}}_n$ and $\mathbf{C}_n$.  We therefore first truncate the singular values and vectors of $\tilde{\mathbf{C}}_n$ to produce $\tilde{\tilde{\mathbf{C}}}_n \in \mathbb{C}^{Q \times 2M}$. Next, we further compress the matrix $\tilde{\tilde{\mathbf{C}}}_n$ into a square matrix $\mathbf{C}_n \in\mathbb{C}^{2M \times 2M}$ which preserves the same singular values and left singular vectors.  This can be done via a process such as that described in \cite{Gordon2019a}.  Following this process, $\mathbf{C}_n$ is square and so can be used for TM recovery as described in Sections \ref{subsec:zeroordermodel} and \ref{subsec:firstordermodel}.

This enables a factorisation of $\tilde{\mathbf{C}}_n$ into the product of $\mathbf{C}_n \in \mathbb{C}^{2M\times 2M}$ and `downsampling matrix', $\mathbf{T}_n \in \mathbb{C}^{P \times 2M}$ such that $\tilde{\mathbf{C}}_n = \mathbf{T}_n \mathbf{C}_n$.  Once the TM has been determined, $\mathbf{T}_n$ can be used to reconstruct $\tilde{\mathbf{C}}_n$ if desired.  The process of downsampling may then be repeated for the other wavelengths, $n=2..4$.

\subsection{Operating process}
\label{subsec:operatingProcessApp}
\begin{figure}[htpb!]
	\centering
	\includegraphics[height=0.7\textheight]{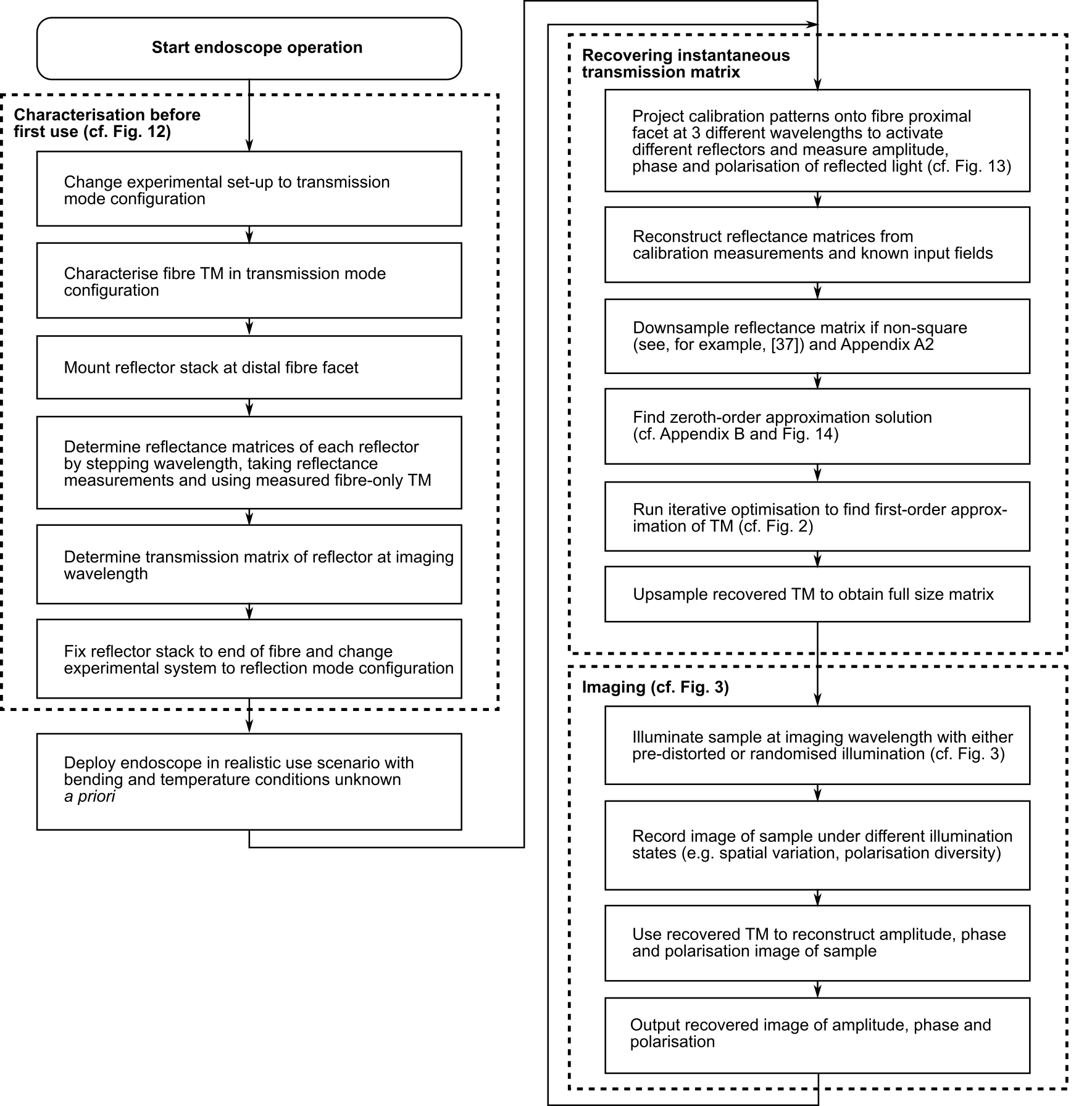}
	\caption{\textbf{Flowchart detailing the full process of TM recovery and imaging using the reflection mode endoscope set-up.}  Details of specific sub-tasks are given in figures and sections as indicated.}
	\label{fig:fullFlowChart}
\end{figure}

Figure \ref{fig:fullFlowChart} shows a flow chart for the full operating process, using the set-up of Figure \ref{fig:proposedExpSetup}, for recovering TMs and performing imaging.

\subsubsection{Characterisation before first use}
\label{subsubsec:charBeforeUseAppendix}
\begin{figure}[htpb]
	\centering
	\includegraphics[height=0.85\textheight]{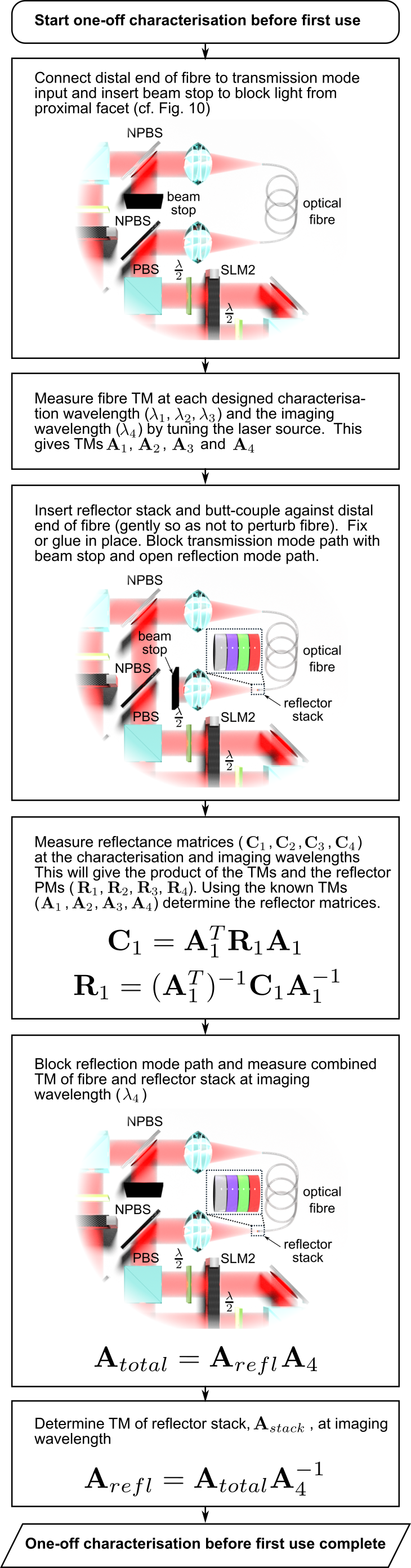}
	\caption{\textbf{Flowchart detailing the series of measurements required to accurately characterise the reflector matrices of the reflector stack at different wavelengths prior to first use of the system.}  This is achieved using the experimental set-up of Figure \ref{fig:proposedExpSetup}.}
	\label{fig:transPreChar}
\end{figure}

Before using the system for the first time, we must take accurate measurements of the reflector stack RMs at all wavelengths and its TM at the imaging wavelength (Figure \ref{fig:transPreChar}).  These are considered to be stable over long periods so this is considered a `one-off' process. A flowchart detailing this process is given in Figure \ref{fig:transPreChar}. First, the fibre-only TMs ($\mathbf{A}_n,~n=1..4$) are characterised using the transmission mode input.  Then, the reflector stack is carefully fixed to the distal fibre facet without perturbing the fibre.  Finally, the characterisation process is repeated to measure the RMs of the fibre with the reflector attached ($\mathbf{C}_n,~n=1..4$). Using the fibre-only TMs, the PMs ($\mathbf{R}_n$ with $n=1..4$) can be determined from the measured RMs. The TM of the reflector stack at the imaging wavelength ($\mathbf{A}_stack$) is also determined at this stage.

Non-square measured TMs $\tilde{\mathbf{A}}_n$ and RMs $\tilde{\mathbf{C}}_n$ with $n=1..4$ may, if needed, be downsampled to form square approximations as described in Appendix \ref{subsec:downsamplingAppendix}.

\subsubsection{Recovering instantaneous transmission matrix}
\label{subsubsec:instTMAppendix}
\begin{figure}[htpb]
	\centering
	\includegraphics[height=0.7\textheight]{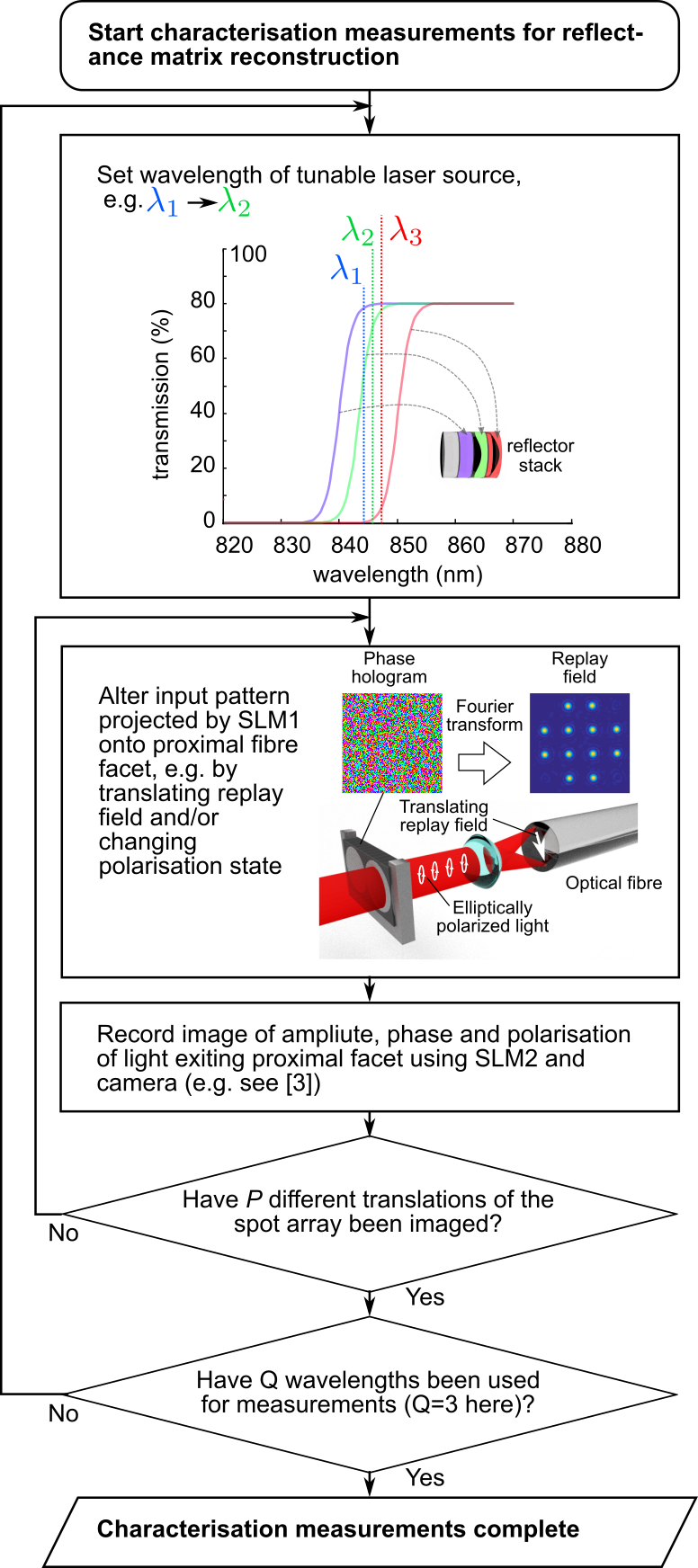}
	\caption{\textbf{Flowchart detailing the series of measurements required to recover instantaneous RMs using the set-up of Figure \ref{fig:proposedExpSetup}.}}
	\label{fig:matMeas}
\end{figure}

The sequence of experimental measurements required to record instantaneous RMs at each respective wavelength ($\mathbf{C}_1$, $\mathbf{C}_2$ and $\mathbf{C}_3$ of Equation \ref{eq:C1} -- Equation \ref{eq:C3}) is illustrated in Figure \ref{fig:matMeas}.  Fundamentally, the process involves sending known calibration patterns into the proximal facet using SLM1 of Figure \ref{fig:proposedExpSetup}, and recording the resultant field after a round-trip (down the fibre, off the distal reflector, back through the fibre) on the camera in the detection arm of Figure \ref{fig:proposedExpSetup}.

Although the RMs at different wavelengths are here measured sequentially, to speed up operation it may be possible to measure them in parallel, owing to linear nature of all optical elements involved. This could, for example, be achieved by using multiple laser sources multiplexed together and using several image sensors, each with its own optical filter, such that light from any given laser source (having a specific wavelength) is detected on only one sensor.

\section{Derivation of zeroth-order model}
\label{subsec:zeroordermodelapp}
Using the zeroth-order assumption implied by Equation \ref{eq:zerothOrderAssumption}, we can rewrite Equations \ref{eq:C1} to \ref{eq:C3} as:

\begin{equation}
\label{eq:C1_zero}
\mathbf{C}_1 = \mathbf{A}^{\top} \mathbf{R}_1 \mathbf{A}
\end{equation}

\begin{equation}
\label{eq:C2_zero}
\mathbf{C}_2 = \mathbf{A}^{\top} \mathbf{R}_2 \mathbf{A}
\end{equation}

\begin{equation}
\label{eq:C3_zero}
\mathbf{C}_3 = \mathbf{A}^{\top} \mathbf{R}_3 \mathbf{A}
\end{equation}

Starting from Equation \ref{eq:C1_zero} and Equation \ref{eq:C2_zero} we derive:

\begin{equation*}
\mathbf{C}_2^{-1} \mathbf{C}_1= \mathbf{A}^{-1} \mathbf{R}_2^{-1} \mathbf{R}_1 \mathbf{A}
\end{equation*}

\noindent which we compactly write as:

\begin{equation}
\label{eq:matrixSim}
\mathbf{C}_{\alpha} = \mathbf{A}^{-1} \mathbf{R}_{\alpha} \mathbf{A}
\end{equation}

\noindent where $\mathbf{C_{\alpha}} = \mathbf{C}_2^{-1} \mathbf{C}_1 \in \mathbb{C}^{2M\times 2M}$, $\mathbf{R}_{\alpha} = \mathbf{R}_2^{-1} \mathbf{R}_1 \in \mathbb{C}^{2M\times 2M}$. Equation \ref{eq:matrixSim} can then be transformed into:

\begin{equation}
\label{eq:sylv1}
\mathbf{R}_{\alpha} \mathbf{A} - \mathbf{A} \mathbf{C}_{\alpha} = \mathbf{0}
\end{equation}

In order to avoid the trivial solution $\mathbf{A} - \mathbf{A} = \mathbf{0}$ to Equation \ref{eq:sylv1}, $\mathbf{R}_{\alpha}$ and $\mathbf{C}_{\alpha}$ must not be equal to the identity matrix, which is ensured by choosing different reflectors $\mathbf{R}_1$ and $\mathbf{R}_2$. It is important to note from Equation \ref{eq:matrixSim} that $\mathbf{C}_{\alpha}$ and $\mathbf{R}_{\alpha}$ are similar matrices and so have the same eigenvalues. In a similar fashion we can derive from Equations \ref{eq:C2_zero} and \ref{eq:C3_zero}:

\begin{equation}
\label{eq:sylv2}
\mathbf{R}_{\beta} \mathbf{A} - \mathbf{A} \mathbf{C}_{\beta} = \mathbf{0}
\end{equation}

\noindent where $\mathbf{C_{\beta}} = \mathbf{C}_3^{-1} \mathbf{C}_2$ and $\mathbf{R}_{\beta} = \mathbf{R}_3^{-1} \mathbf{R}_2$.  Finally, from  Equations \ref{eq:C1_zero} and \ref{eq:C3_zero}:

\begin{equation}
\label{eq:sylv3}
\mathbf{R}_{\gamma} \mathbf{A} - \mathbf{A} \mathbf{C}_{\gamma} = \mathbf{0}
\end{equation}

\noindent where $\mathbf{C_{\gamma}} = \mathbf{C}_3^{-1} \mathbf{C}_1$ and $\mathbf{R}_{\gamma} = \mathbf{R}_3^{-1} \mathbf{R}_1$. Equations \ref{eq:sylv1}--\ref{eq:sylv3} are examples of Sylvester equations so we can exploit existing methods to solve them. Because Equations \ref{eq:sylv1}--\ref{eq:sylv3} have RHS $=\mathbf{0}$, the trivial solution $\mathbf{A} = \mathbf{0}$ satisifies all three.  However, we know that there must exist at least one non-trivial solution for $\mathbf{A}$ because if $\mathbf{A} = \mathbf{0}$ then $\mathbf{C}_1, \mathbf{C}_2, \mathbf{C}_3   = \mathbf{0}$ by Equations \ref{eq:C1}--\ref{eq:C3}. This would imply a system with 100\% power loss which we know physically is not the case. Therefore, we aim to determine the space of non-trivial solutions (i.e. the nullspace of Equation \ref{eq:sylv1}, \ref{eq:sylv2} or \ref{eq:sylv3}) by implementing the Bartels--Stewart algorithm for solving Sylvester equations of the form $\mathbf{D}\mathbf{X} - \mathbf{X}\mathbf{E} = \mathbf{F}$ for the special case where $\mathbf{F} = \mathbf{0}$ \cite{Bartels1972}.

We begin by using the Schur matrix decomposition \cite{Horn2013}.  This states that any square matrix, $\mathbf{H}$ can be decomposed as:

\begin{equation}
\mathbf{H} = \mathbf{Q} \mathbf{U} \mathbf{Q}^{-1}
\end{equation}

\noindent where $\mathbf{Q}$ is unitary and $\mathbf{U}$ is an upper triangular matrix. Importantly, the diagonal of $\mathbf{U}$ comprises the eigenvalues of $\mathbf{H}$ sorted in descending order of magnitude. With reference to Equation \ref{eq:sylv1} we can then write:

\begin{equation}
\label{eq:Rschur}
\mathbf{R}_{\alpha} = \mathbf{Q}_{R_\alpha} \mathbf{U}_{R_\alpha} \mathbf{Q}_{R_\alpha}^{-1}
\end{equation}

\begin{equation}
\label{eq:Cschur}
\mathbf{C}_{\alpha} = \mathbf{Q}_{C_\alpha} \mathbf{U}_{C_\alpha} \mathbf{Q}_{C_\alpha}^{-1}
\end{equation}

Substituting Equation \ref{eq:Rschur} and Equation \ref{eq:Cschur} back into Equation \ref{eq:sylv1} gives:

\begin{equation*}
\mathbf{Q}_{R_\alpha} \mathbf{U}_{R_\alpha} \mathbf{Q}_{R_\alpha}^{-1} \mathbf{A} - \mathbf{A} \mathbf{Q}_{C_\alpha} \mathbf{U}_{C_\alpha} \mathbf{Q}_{C_\alpha}^{-1}  = \mathbf{0}
\end{equation*}

\begin{equation*}
\mathbf{U}_{R_\alpha} \mathbf{Q}_{R_\alpha}^{-1} \mathbf{A} \mathbf{Q}_{C_\alpha} - \mathbf{Q}_{R_\alpha} ^{-1} \mathbf{A} \mathbf{Q}_{C_\alpha} \mathbf{U}_{C_\alpha}  = \mathbf{0}
\end{equation*} 

\begin{equation}
\label{eq:schurSylv}
\mathbf{U}_{R_\alpha} \mathbf{A}' - \mathbf{A}' \mathbf{U}_{C_\alpha}  = \mathbf{0}
\end{equation}

\noindent where 

\begin{equation}
\label{eq:gettingAdashFromA}
\mathbf{A}' = \mathbf{Q}_{R_\alpha}^{-1} \mathbf{A} \mathbf{Q}_{C_\alpha}
\end{equation}

\noindent and $\mathbf{A}' \in \mathbb{C}^{2M \times 2M}$. Because of the degenerate nature of Equation \ref{eq:schurSylv} (i.e. RHS $=\mathbf{0}$) there are multiple solutions to this equation  -- the trivial solution plus at least one non-trivial.  We reduce the space of possible solutions by requiring that the eigenvalues of $\mathbf{R}_{\alpha}$ are distinct, with the result that only the diagonal elements of $\mathbf{A}'$ are indeterminate (see Appendix \ref{subsec:appendixSolvingSylv} for full derivation). This is arranged by suitable design of the reflectors $\mathbf{R}_1$ and $\mathbf{R}_2$, which is not in general difficult because even randomly generated matrices produce distinct eigenvalues with high probability \cite{Marcenko1967,Popoff2010}. 

Equation \ref{eq:schurSylv} is another Sylvester equation but, crucially, $\mathbf{U}_{R_\alpha}$ and $\mathbf{U}_{C_\alpha}$ are upper triangular matrices so we can solve for $\mathbf{A}'$ element by element as it is typically done in the Bartels--Stewart algorithm. It follows that elements of $\mathbf{A}'$ are related to elements of $\mathbf{U}_{C_\alpha}$ and $\mathbf{U}_{R_\alpha}$ as:

\begin{equation}
\label{eq:schurSolution}
\begin{array}{c}
\left[ur_{P,Q} a_{Q,Q} - \sum_{s=P}^{Q-1} uc_{s,Q} a_{P,s}\right] + (ur_{P,P} - uc_{Q,Q})  a_{P,Q} + \\
 \sum_{q=P+1}^{Q-1} ur_{P,q} a_{q,Q} = 0
\end{array}
\end{equation}

\noindent where $uc_{P,Q}$ represents an element of $\mathbf{U}_{C_\alpha}$ at row $P$ and column $Q$, $ur_{P,Q}$ represents an element of $\mathbf{U}_{R_\alpha}$ at row $P$ column $Q$, and $a_{P,Q}$ represents an element of $\mathbf{A}'$ at row $P$ and column $Q$ (see Appendix \ref{subsec:appendixSolvingSylv} for full derivation). 

Varying $P$ from $1$ to $Q-1$ in Equation \ref{eq:schurSolution} produces $Q-1$ equations.  If we arbitrarily set the diagonal elements, $a_{m,m}$, of $\mathbf{A'}$, we are left with $Q-1$ unknowns and can solve this system of linear equations for all elements of the $Q^{th}$ column of $\mathbf{A'}$.  By varying $Q$ from $2$ to $2M$ and solving each resultant system of equations, we can fully determine all non-diagonal elements of $\mathbf{A}'$ given a set of arbitrary diagonal elements, $a_{m,m}$ ($m=1.. 2M$).  This repeated solution of systems of linear equations could be expressed as a single matrix equation, but the matrix would be of size $M(2M-1) \times M(2M-1)$ creating a memory requirement $\sim M^4$.  The equation systems are therefore solved sequentially for computational tractability.

\begin{figure}[htbp]
	\centering
	\includegraphics[width=\linewidth]{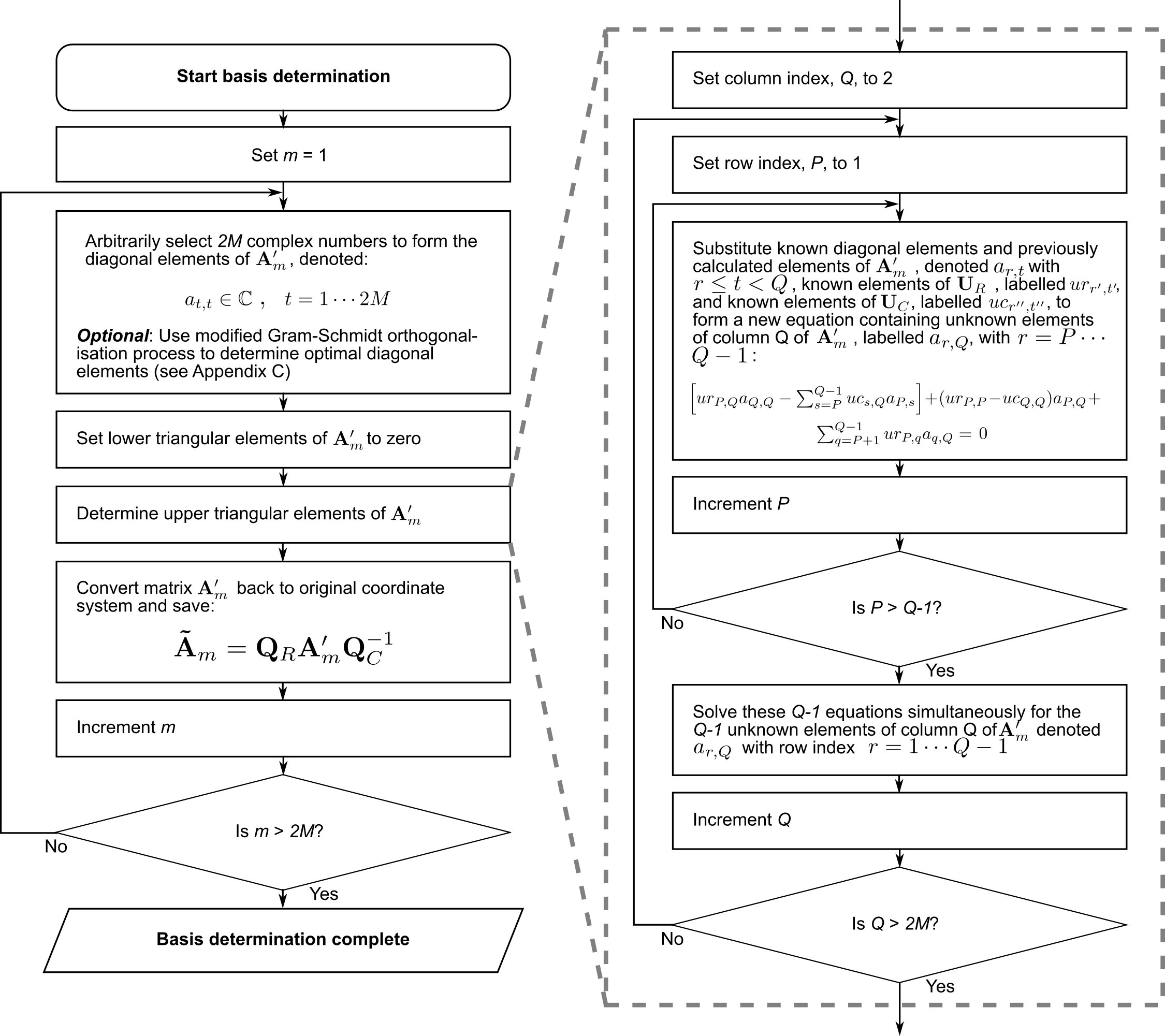}
	\caption{\textbf{Flowchart detailing the process of finding a basis for the matrix $\mathbf{A}$ using the zeroth-order solution method.}}
	\label{fig:zerothOrderFlowchart}
\end{figure}

When an arbitrary set of $2M$ complex numbers (i.e. a vector $\in \mathbb{C}^{2M}$) is selected for the diagonal elements of $\mathbf{A}'$ in Equation \ref{eq:schurSylv} (or equivalently elements $a_{m,m}$ where $m=1.. 2M$ of Equation \ref{eq:schurSolution}) the remaining elements of $\mathbf{A}'$ can be solved. This implies that the `true' solution for $\mathbf{A}'$ can be expressed as a linear combination of at most $2M$ matrices, $\mathbf{A}_m'$ where $m=1 .. 2M$, each of which is a solution of Equation \ref{eq:schurSylv}.  Together, these $2M$ matrices comprise a basis for $\mathbf{A}'$ (equivalently, they comprise a nullspace of Equation \ref{eq:schurSylv}). Using Equation \ref{eq:gettingAdashFromA} this basis can be converted to a basis for $\mathbf{A}$ (equivalently, a nullspace for Equation \ref{eq:sylv1}), denoted $\mathbf{\tilde{A}}_m$ where $m=1.. 2M$. To improve robustness to noise and numerical error an iterative Gram--Schmidt orthogonalisation procedure, modified to account for the conversion from $\mathbf{A'}_m$ to $\mathbf{\tilde{A}}_m$, can be applied to create an orthogonal set of $\mathbf{\tilde{A}}_m$ where $m=1 .. 2M$ (see Appendix \ref{sec:orthogonalisingBasisAppendix}). The process of determining a basis for $\mathbf{A}$ is summarised in Figure \ref{fig:zerothOrderFlowchart}. 

Having determined a suitable basis we express the matrix we wish to recover, $\mathbf{A}$, as:

\begin{equation}
\label{eq:reducedSolution}
\mathbf{A} = w_1 \mathbf{\tilde{A}}_1 + .. + w_{2M} \mathbf{\tilde{A}}_{2M}
\end{equation}

\noindent where $w_m$ represents the complex-valued weight of the $m^{th}$ basis element, the matrix $\mathbf{\tilde{A}}_m$. The solution space now has only $2M$ degrees of freedom (represented by $w_m$) so computational complexity can be further reduced by considering a subset of elements of $\mathbf{A}$. We can select $B$ ($\geq 2M$) matrix elements of $\mathbf{A}$, either arbitrarily or according to some prior information e.g. the elements with the largest mean based on a statistical model.  The corresponding $B$ elements of $\mathbf{\tilde{A}}_m$ are then ordered into a column vector $\mathbf{b}_m \in \mathbb{C}^B$ for every $m=1 .. 2M$. These column vectors form a matrix $\mathbf{B}_\alpha \in \mathbb{C}^{B \times 2M}$:

\begin{equation}
\mathbf{B}_\alpha = \left[\begin{array}{c c c c}
\mathbf{b}_1 & \mathbf{b}_2  & \cdots & \mathbf{b}_{2M}
\end{array}\right]
\end{equation}

\noindent so that we can write:

\begin{equation}
\label{eq:firstB_est}
\mathbf{B}_\alpha \mathbf{w}^{\top} = \mathbf{b}_{est}
\end{equation}

\noindent where $\mathbf{b}_{est}$ is an estimate of the $B$ selected elements of the true TM, $\mathbf{A}$, and $\mathbf{w} = [w_1~~\cdots ~~w_N]$ is a vector containing the complex weights of Equation \ref{eq:reducedSolution}. Since $\mathbf{B}_\alpha$ is either a square or tall matrix, we pre-multiply by its Moore-Penrose pseudo inverse, $\mathbf{B}_\alpha^{\dagger}$:

\begin{equation*}
\mathbf{B}_\alpha^{\dagger} \mathbf{B}_\alpha \mathbf{w}^{\top} = \mathbf{B}_\alpha^{\dagger} \mathbf{b}_{est}
\end{equation*}

\begin{equation*}
\mathbf{w}^{\top} = \mathbf{B}_\alpha^{\dagger} \mathbf{b}_{est}
\end{equation*}

We then multiply both sides by $\mathbf{B}_\alpha$ to get:

\begin{equation*}
\mathbf{B}_\alpha \mathbf{w}^{\top} = \mathbf{B}_\alpha \mathbf{B}_\alpha^{\dagger} \mathbf{b}_{est}
\end{equation*}

\noindent and substitute in Equation \ref{eq:firstB_est} to obtain a recursive expression:

\begin{equation}
\label{eq:directReducedSpace}
\mathbf{b}_{est} = \mathbf{B}_\alpha \mathbf{B}_\alpha^{\dagger} \mathbf{b}_{est}
\end{equation}

Following the same steps to derive Equation \ref{eq:directReducedSpace} from Equation \ref{eq:sylv1} but starting from Equation \ref{eq:sylv2} gives:

\begin{equation}
\label{eq:betaEq}
\mathbf{b}_{est} = \mathbf{B}_{\beta} \mathbf{B}_{\beta}^{\dagger} \mathbf{b}_{est}
\end{equation}

Similarly, starting from Equation \ref{eq:sylv3} gives:

\begin{equation}
\label{eq:gammaEq}
\mathbf{b}_{est} = \mathbf{B}_{\gamma} \mathbf{B}_{\gamma}^{\dagger} \mathbf{b}_{est}
\end{equation}

Clearly, $\mathbf{b}_{est}$ is an eigenvector of $\mathbf{B}_{\alpha} \mathbf{B}_{\alpha}^{\dagger}$, $\mathbf{B}_{\beta} \mathbf{B}_{\beta}^{\dagger}$ and $\mathbf{B}_{\gamma} \mathbf{B}_{\gamma}^{\dagger}$ with an eigenvalue of 1. Physical considerations of power conservation suggest the existence of at least one non-trivial solution for $\mathbf{b}_{est}$.  Our empirical investigations with both simulated and real TMs (see Section \ref{sec:results}) suggest that when using at least two out of Equations \ref{eq:directReducedSpace}, \ref{eq:betaEq} and \ref{eq:gammaEq} we can identify a single non-trivial solution, the `true' solution: we arbitrarily choose Equations \ref{eq:directReducedSpace} and \ref{eq:betaEq}.  Next we apply a variant of the `power method' for finding dominant eigenvalues of a matrix \cite{Bodewig1959} and recursively substitute Equation \ref{eq:directReducedSpace} into Equation \ref{eq:betaEq}, substitute the result into Equation \ref{eq:directReducedSpace} etc. to give:

\begin{equation}
\mathbf{b}_{est}^t = \mathbf{B}_{\alpha} \mathbf{B}_{\alpha}^{\dagger} \mathbf{B}_{\beta} \mathbf{B}_{\beta}^{\dagger} \dots \mathbf{B}_{\alpha} \mathbf{B}_{\alpha}^{\dagger} \mathbf{b}_{est}^0
\end{equation}

\noindent where $t$ denotes the number of matrix multiplications.  For large $t$, the output $\mathbf{b}_{est}^t$ is expected to converge to the desired solution $\mathbf{b}_{est}$ for any input $\mathbf{b}_{est}^0$. Alternatively, for large $t$, the dominant eigenvector of $\mathbf{B}_{\alpha} \mathbf{B}_{\alpha}^{\dagger} \mathbf{B}_{\beta} \mathbf{B}_{\beta}^{\dagger} \dots \mathbf{B}_{\alpha} \mathbf{B}_{\alpha}^{\dagger}$ will approximate the solution well. In this work, we find empirically that 2--3 iterations, i.e.~$t\in\{4,6\}$, is sufficient for faithful TM recovery.

To obtain the desired full TM, $\mathbf{A}$, we determine the weights, $\mathbf{w}$, from approximation $\mathbf{b}_{est}^t$ using Equation \ref{eq:firstB_est} and then compute the appropriate weighted sum of $\mathbf{A}_m$ with $m=1 .. 2M$ using Equation \ref{eq:reducedSolution}.

An alternative to such a method could be to directly optimise the weights of Equation \ref{eq:reducedSolution} to best satisfy Equations \ref{eq:C1_zero}--\ref{eq:C3_zero}, or to minimise a prior probability distribution placed on the matrix (e.g. an empirically determined sparsity structure such as that presented in \cite{Gordon2019a}).

\subsection{Sylvester equations element by element using Bartels-Stewart algorithm}
\label{subsec:appendixSolvingSylv}
The Bartels-Stewart algorithm is a method of solving matrix equations of the form $\mathbf{D}\mathbf{X} - \mathbf{X}\mathbf{E} = \mathbf{F}$ (where $\mathbf{D}$, $\mathbf{E}$, $\mathbf{F}$ $\in \mathbb{C}^{N \times N}$), termed \emph{Sylvester equations}, to find the matrix $\mathbf{X} \in \mathbb{C}^{N \times N}$ \cite{Bartels1972}.  Here, we wish to solve the special case where $\mathbf{F} = \mathbf{0}$. We begin with the Schur decompositions of the matrices $\mathbf{R}_{\alpha}$ and $\mathbf{C}_{\alpha}$ (adapted from Equations \ref{eq:Rschur} and \ref{eq:Cschur}):

\begin{equation}
\label{eq:RschurApp}
\mathbf{R} = \mathbf{Q}_R \mathbf{U}_R \mathbf{Q}_R^{-1}
\end{equation}

\begin{equation}
\label{eq:CschurApp}
\mathbf{C} = \mathbf{Q}_C \mathbf{U}_C \mathbf{Q}_C^{-1}
\end{equation}

\noindent where $\mathbf{Q}_R$ and $\mathbf{Q}_R$ is unitary and $\mathbf{U}_R$ and $\mathbf{U}_C$ are upper triangular matrices. Importantly, the diagonal of $\mathbf{U}_C$ comprises the eigenvalues of $\mathbf{C}$ sorted in descending order of magnitude, and likewise for $\mathbf{U}_R$. The Schur decomposition of a matrix with entries sorted in this way can be computed, for example, in MATLAB.  This enables us to write the following Sylvester equation:

\begin{equation}
\label{eq:schurSylvApp}
\mathbf{U}_R \mathbf{A}' - \mathbf{A}' \mathbf{U}_C  = \mathbf{0}
\end{equation}

\noindent where $\mathbf{A}'$ is linked to the original transmission matrix, $\mathbf{A}$ by:

\begin{equation}
\label{eq:gettingAdashFromAApp}
\mathbf{A}' = \mathbf{Q}_R^{-1} \mathbf{A} \mathbf{Q}_C
\end{equation}

Recalling that all matrices here are $\in \mathbb{C}^{N \times N}$, we first consider the element in row $N$, column $1$, i.e. $(N,1)$, of the zero matrix on the RHS of Equation \ref{eq:schurSylvApp}.  We see that:

\begin{equation}
\label{eq:elementWise1App}
ur_{N,N}  a_{N,1} - a_{N,1} uc_{1,1} = 0
\end{equation}

\noindent where $ur_{m,n}$ is the element in row $m$, column $n$ of $\mathbf{U}_R$, $uc_{m,n}$ is the element in row $m$, column $n$ of $\mathbf{U}_C$, and $a_{m,n}$ is the element in row $m$, column $n$ of $\mathbf{A}'$.

Next, we require that the eigenvalues of $\mathbf{R}$ are distinct, which is ensured by suitable design of reflectors (see Sections \ref{subsec:physicalModel} and \ref{subsec:implementingReflectors}). We also know that the eigenvalues of $\mathbf{R}$ are equal to those of $\mathbf{C}_{\alpha}$ because the two are similar matrices (see, for example, Equation \ref{eq:matrixSim}). Therefore, the values on the diagonal of $\mathbf{U}_R$ are distinct from one another, and are equal to the values on the diagonal of $\mathbf{U}_C$.  The only way that Equation \ref{eq:elementWise1App} can hold is if $a_{N,1} = 0$, providing the solution for that element.

We then consider element $(N-1,1)$, giving: 

\begin{equation*}
ur_{N-1,N-1}  a_{N-1,1} + ur_{N-1,N}  a_{N,1} - a_{N-1,1} uc_{1,1} = 0
\end{equation*}

Since we know that $a_{N,1} = 0$, we apply again the above reasoning and conclude that $a_{N-1,1} = 0$.  Continuing on up this column, we find that all elements must be zero until we reach the first row, where:

\begin{equation*}
ur_{1,1}  a_{1,1} - a_{1,1} uc_{1,1} = 0
\end{equation*}

Now, we know that $ur_{m,m} = uc_{m,m}$ for every $m$ as discussed above.  Therefore, every possible $a_{1,1} \in \mathbb{C}$ satisfies this equation, meaning the element is indeterminate. We now consider the second column of the zero matrix on the RHS of Equation \ref{eq:schurSylvApp}, starting with element  $(N,2)$:

\begin{equation*}
ur_{N,N}  a_{N,2} - (a_{N,1} uc_{1,2} + a_{N,2} uc_{2,2}) = 0
\end{equation*}

We know $a_{N,1} = 0$ so can write:

\begin{equation*}
ur_{N,N}  a_{N,2} - a_{N,2} uc_{2,2} = 0
\end{equation*}

\noindent and therefore $a_{N,2} = 0$. Next, we consider element $(N-1,2)$:

\begin{equation*}
ur_{N-1,N-1}  a_{N-1,2} + ur_{N-1,N} a_{N,2} - (a_{N-1,1} uc_{1,2} + a_{N-1,2} uc_{2,2}) = 0
\end{equation*}

Using previously known elements, we find that:

\begin{equation*}
ur_{N-1,N-1}  a_{N-1,2} - a_{N-1,2} uc_{2,2} = 0
\end{equation*}

Therefore $a_{N-1,2} = 0$.  This continues up column 2 of the zero matrix until we get to:

\begin{equation*}
ur_{2,2}  a_{2,2} - a_{2,2} uc_{2,2} = 0
\end{equation*}

$ a_{2,2}$ can be anything since $ur_{2,2}  = uc_{2,2}$.  Now, we consider element $(1,2)$:

\begin{equation*}
ur_{1,1}  a_{1,2} + ur_{1,2} a_{2,2} - (a_{1,1} uc_{1,2} + a_{1,2} uc_{2,2}) = 0
\end{equation*}

\begin{equation*}
(ur_{1,1} - uc_{2,2})  a_{1,2} + ur_{1,2} a_{2,2} - a_{1,1} uc_{1,2} = 0
\end{equation*}

By repeatedly applying this logic to each column of the zero matrix on the RHS of Equation \ref{eq:schurSylvApp}, we conclude that the matrix $\mathbf{A}'$ must have an upper triangular form. It then follows that element $(3,3)$ gives:

\begin{equation*}
ur_{3,3}  a_{3,3} - a_{3,3} uc_{3,3} = 0
\end{equation*}

and element $(2,3)$ gives:

\begin{equation*}
ur_{2,1}  a_{1,3} + ur_{2,2}  a_{2,3} + ur_{2,3}  a_{3,3} - (a_{2,1} uc_{1,3} + a_{2,2} uc_{2,3} +  a_{2,3} uc_{3,3} ) = 0
\end{equation*}

\begin{equation*}
ur_{2,1}  a_{1,3} + (ur_{2,2} - uc_{3,3})  a_{2,3} + ur_{2,3}  a_{3,3} - a_{2,2} uc_{2,3}  = 0
\end{equation*}

Considering element $(1,3)$ gives:

\begin{equation*}
ur_{1,1}  a_{1,3} + ur_{1,2}  a_{2,3} + ur_{1,3}  a_{3,3} - (a_{1,1} uc_{1,3} + a_{1,2} uc_{2,3} +  a_{1,3} uc_{3,3} ) = 0
\end{equation*}

\begin{equation*}
(ur_{1,1} - uc_{3,3})  a_{1,3} + ur_{1,2}  a_{2,3} + ur_{1,3}  a_{3,3} - a_{1,1} uc_{1,3} - a_{1,2} uc_{2,3} = 0
\end{equation*}

Therefore, these elements are all functions of elements that have already been computed when solving previously columns of the zero matrix on the RHS of Equation \ref{eq:schurSylvApp}, and the indeterminate diagonal elements of $\mathbf{A}'$.  By repeated application of this process, we can generate an equation for each upper triangular element, $(P,Q)$ with $P < Q$, of the zero matrix on the RHS of Equation \ref{eq:schurSylvApp} as:

\begin{equation*}
(ur_{P,P} - uc_{Q,Q})  a_{P,Q} + \sum_{q=P+1}^{Q} ur_{P,q} a_{q,Q} - \sum_{s=P}^{Q-1} uc_{s,Q} a_{P,s} = 0
\end{equation*}

Rearranging, we can write:

\begin{equation*}
\left[- \sum_{s=P}^{Q-1} uc_{s,Q} a_{P,s}\right] + (ur_{P,P} - uc_{Q,Q})  a_{P,Q} + \sum_{q=P+1}^{Q} ur_{P,q} a_{q,Q} = 0
\end{equation*}

\begin{equation}
\label{eq:schurSolutionApp}
\left[ur_{P,Q} a_{Q,Q} - \sum_{s=P}^{Q-1} uc_{s,Q} a_{P,s}\right] + (ur_{P,P} - uc_{Q,Q})  a_{P,Q} + \sum_{q=P+1}^{Q-1} ur_{P,q} a_{q,Q} = 0
\end{equation}

In summary, we find that the diagonal elements, $a_{m,m}$ with $m=1..N$, are indeterminate but that all other elements in the matrix $\mathbf{A}'$ can be computed from these diagonals by solving a series of linear equations.

\section{Orthogonal basis generation}
\label{sec:orthogonalisingBasisAppendix}
To achieve maximum signal-to-noise ratio during optimisation or any other solution method, the basis generated for reconstructing the TM, $\mathbf{A}$ in Section \ref{subsec:zeroordermodel}, should ideally be orthogonal. This basis is constructed starting from an arbitrary vector $\in \mathbb{C}^{N}$ used for the diagonal elements (Figure \ref{fig:zerothOrderFlowchart}). In general, this will produce a complete but non-orthogonal basis.  Elements of this basis may be very similar resulting in numerical instabilities during the zeroth-order recovery process.  

To produce an orthogonal basis, we first define a function that gives the solution of the system of equations defined in Equation \ref{eq:schurSolution}. This function, which could be implemented by any linear equation solver, takes as input a vector representing diagonal elements of the solution matrix, $\mathbf{u}$, and returns the full matrix, $\mathbf{B}$:

\begin{equation}
\mathbf{B} = f(\mathbf{u})
\end{equation}

We can then start with an arbitrary set of complex diagonal elements, $\mathbf{u}_1 \in \mathbb{C}^N$, to get a basis element for $\mathbf{A}'$ (defined in Equation \ref{eq:gettingAdashFromAApp}), which we term $\mathbf{B}_1' = f(\mathbf{u}_1)$. Next, we invert Equation \ref{eq:gettingAdashFromAApp} to find the associated basis element for $\mathbf{A}$, given by $\mathbf{B}_1 = \mathbf{Q}_R \mathbf{B}_1' \mathbf{Q}_C^{-1}$. To get the next basis element we then generate a random complex matrix, $\mathbf{B}_2 \in \mathbb{C}^{N\times N}$ and perform a Gram-Schmidt orthogonalisation step:

\begin{equation}
\label{eq:gramSchmidtStep}
\mathrm{vec} (\mathbf{\hat{B}}_2) = \mathrm{vec}(\mathbf{B}_2) - \mathrm{proj}_{\mathrm{vec} (\mathbf{B}_1)} \mathrm{vec}(\mathbf{B}_2)
\end{equation}

\noindent where $\mathrm{vec}(..)$ represents ordering of matrix elements into a vector. We then use Equation \ref{eq:gettingAdashFromAApp} to convert back to an estimated basis element for $\mathbf{A}'$:

\begin{equation}
\mathbf{\hat{B}}_2' = \mathbf{Q}_R^{-1} \mathbf{\hat{B}}_2 \mathbf{Q}_C
\end{equation}

\noindent and find the diagonal elements of this basis element:

\begin{equation}
\mathbf{\hat{u}}_2 = \mathrm{diag} \left(\mathbf{\hat{B}}_2'\right)
\end{equation}

We then use these diagonal elements to generate a new matrix that is a valid solution to \ref{eq:schurSolution} (i.e. a basis element for $\mathbf{A}'$):

\begin{equation}
\mathbf{B}_2' = f(\mathbf{\hat{u}}_2)
\end{equation}

\noindent and convert back to a basis element for $\mathbf{A}$:

\begin{equation}
\mathbf{B}_2 = \mathbf{Q}_R \mathbf{B}_2' \mathbf{Q}_C^{-1}
\end{equation}

We then substitute $\mathbf{B}_2$ back into Equation \ref{eq:gramSchmidtStep} and iterate several times.  The result of this will be a basis matrix, $\mathbf{B}_2$, that is orthogonal to $\mathbf{B}_1$. This process is then repeated for subsequent basis elements using all previously generated basis elements.  For example, the Gram-Schmidt orthogonalisation procedure step for the $m^{th}$ basis element becomes:

\begin{equation}
\mathrm{vec} (\mathbf{\hat{B}}_m) = \mathrm{vec}(\mathbf{B}_m) - \sum_{j=1}^m \mathrm{proj}_{\mathrm{vec} (\mathbf{B}_j)} \mathrm{vec}(\mathbf{B}_m)
\end{equation}

In this way, an orthogonal basis for $\mathbf{A}$ can be generated.

\section{Alternative method of finding the null space of Sylvester equations}
In some cases, finding a basis for the TM can suffer from numerical error due to the large number of computation steps involved (see Section \ref{subsec:zeroordermodel}). To derive an alternative approach, we begin with Equation \ref{eq:sylv1}:

\begin{equation}
\label{eq:sylv1App}
\mathbf{R}_{\alpha} \mathbf{A} - \mathbf{A} \mathbf{C}_{\alpha} = \mathbf{0}
\end{equation}

We can rewrite this in the following form \cite{Horn2013}:

\begin{equation}
\mathbf{M}~~\mathrm{vec} (\mathbf{A}) = 0
\end{equation}

\noindent where $\mathrm{vec}(..)$ represents ordering of matrix elements into a vector and

\begin{equation}
\mathbf{M} = \mathbf{I}_{N \times N} \otimes \mathbf{R}_{\alpha} + \mathbf{R}_{\alpha} \otimes \mathbf{I}_{N \times N}
\end{equation}

\noindent where $\otimes$ is the Kronecker product.  By finding the nullspace of $\mathbf{M}$ we find an orthogonal basis for $\mathbf{A}$. If $\mathbf{A}$ is of size $N \times N$, the resultant matrix $\mathbf{M}$ is of size $N^2 \times N^2$ and so for larger problems (e.g. the 1648$\times$1648 MCF TMs of Section \ref{subsec:realMCFresults}), the memory usage becomes impractically large.  Therefore, in this work this method is used for small simulated matrices.

\section{Approximate bandwidth calculation}
\label{sec:bwCalcAppendix}
\noindent Let us call the centre wavelength of the fibre $\lambda_0$ and the bandwidth of interest $\Delta \lambda$. The bandwidth in frequency units is:

\begin{equation*}
f_1 - f_2 = \frac{c}{n_{eff} (\lambda_0 - \Delta \lambda/2)} - \frac{c}{n_{eff} (\lambda_0 + \Delta \lambda/2)}
\end{equation*}

\begin{equation}
\Delta f= \frac{c}{n_{eff}} \left(\frac{1}{\lambda_0 - \Delta \lambda/2} - \frac{1}{\lambda_0 + \Delta \lambda/2}\right)
\end{equation}

\noindent where $n_{eff}$ is the effective refractive index of the fibe (e.g. $\sim$1.5 for glass) and $c$ is the speed of light in a vacuum. This bandwidth is the range over which the eigenmodes of the fibre remain constant, which is equivalent to the spectral correlation bandwidth \cite{Mosk2012}.  This is in turn related to the time difference between longest path lengths (time of flight) by:

\begin{equation}
\Delta t = \frac{1}{\Delta f} = \frac{n_{eff}}{c} \frac{1}{\left(\frac{1}{\lambda_0 - \Delta \lambda/2} - \frac{1}{\lambda_0 + \Delta \lambda/2}\right)}
\end{equation}

If we assume the fibre has a physical length $\ell$, we find that time of flight down the fibre for the fastest mode will be of the order of:

\begin{equation}
t_1 = \frac{\ell}{\frac{c}{n_{eff}}}
\end{equation}

The longest time of flight will then be:

\begin{equation}
t_2 = t_1 + \Delta t
\end{equation}

This results in an effective length for the longest path of:

\begin{equation}
\ell_2 = \frac{c}{n_{eff}} t_2 = \frac{c}{n_{eff}} \left(t_1 + \Delta t\right)
\end{equation}

\begin{equation}
\ell_2 = \frac{c}{n_{eff}} \left(\frac{\ell}{\frac{c}{n_{eff}}} + \Delta t\right) = \ell + \frac{c}{n_{eff}} \Delta t
\end{equation}

\begin{equation}
\ell_2 =  \ell + \frac{1}{\left(\frac{1}{\lambda_0 - \Delta \lambda/2} - \frac{1}{\lambda_0 + \Delta \lambda/2}\right)}
\end{equation}

At the shortest wavelength, $\lambda_0 - \Delta \lambda/2$, the phase shift introduced by the shortest path is:

\begin{equation}
\phi_{1,short} = \frac{2\pi \ell}{\lambda_0 - \Delta \lambda/2}
\end{equation}

\noindent and by the longest path:

\begin{equation}
\phi_{1,long} = \frac{2\pi \ell_2}{\lambda_0 - \Delta \lambda/2}
\end{equation}

\begin{equation}
\phi_{1,long} = \frac{2\pi\left( \ell + \frac{1}{\left(\frac{1}{\lambda_0 - \Delta \lambda/2} - \frac{1}{\lambda_0 + \Delta \lambda/2}\right)}\right)}{\lambda_0 - \Delta \lambda/2}
\end{equation}

\noindent giving a difference of:

\begin{equation}
\Delta \phi_1 = \phi_{1,long} - \phi_{1,short} = 2\pi \frac{\frac{1}{\left(\frac{1}{\lambda_0 - \Delta \lambda/2} - \frac{1}{\lambda_0 + \Delta \lambda/2}\right)}}{\lambda_0 - \Delta \lambda/2}
\end{equation}

\begin{equation}
\Delta \phi_1 = 2\pi \frac{1}{\left(1 - \frac{\lambda_0 - \Delta \lambda/2}{\lambda_0 + \Delta \lambda/2}\right)}
\end{equation}

At the longest wavelength, $\lambda_0 + \Delta \lambda/2$, the phase shift introduced by the shortest path is:

\begin{equation}
\phi_{2,short} = \frac{2\pi \ell}{\lambda_0 + \Delta \lambda/2}
\end{equation}

\noindent and by the longest path:

\begin{equation}
\phi_{2,long} = \frac{2\pi \ell_2}{\lambda_0 + \Delta \lambda/2}
\end{equation}

\begin{equation}
\phi_{2,long} = \frac{2\pi\left( \ell + \frac{1}{\left(\frac{1}{\lambda_0 - \Delta \lambda/2} - \frac{1}{\lambda_0 + \Delta \lambda/2}\right)}\right)}{\lambda_0 + \Delta \lambda/2}
\end{equation}

\noindent giving a difference of:

\begin{equation}
\Delta \phi_2 = \phi_{2,long} - \phi_{2,short} = 2\pi \frac{\frac{1}{\left(\frac{1}{\lambda_0 - \Delta \lambda/2} - \frac{1}{\lambda_0 + \Delta \lambda/2}\right)}}{\lambda_0 + \Delta \lambda/2}
\end{equation}

\begin{equation}
\Delta \phi_2 = 2\pi \frac{1}{\left(\frac{\lambda_0 + \Delta \lambda/2}{\lambda_0 - \Delta \lambda/2}- 1\right)}
\end{equation}

Now we consider the difference in phase shift between the longest and shortest path at the two wavelengths, which corresponds to the phase difference between respective elements of matrices at the two different wavelengths.  This gives:

\begin{equation}
\Delta \phi_1 - \Delta \phi_2 = 2\pi \left(\frac{1}{\left(1 - \frac{\lambda_0 - \Delta \lambda/2}{\lambda_0 + \Delta \lambda/2}\right)} -  \frac{1}{\left(\frac{\lambda_0 + \Delta \lambda/2}{\lambda_0 - \Delta \lambda/2}- 1\right)}\right)
\end{equation}

\begin{equation}
\Delta \phi_1 - \Delta \phi_2 = 2\pi \left(\frac{1}{\left(1 - \frac{\lambda_0 - \Delta \lambda/2}{\lambda_0 + \Delta \lambda/2}\right)} +  \frac{1}{\left(1 - \frac{\lambda_0 + \Delta \lambda/2}{\lambda_0 - \Delta \lambda/2}\right)}\right)
\end{equation}

\begin{equation}
\Delta \phi_1 - \Delta \phi_2 = 2\pi \left(\frac{\lambda_0 + \Delta \lambda/2}{\left(\lambda_0 + \Delta \lambda/2 - \lambda_0 + \Delta \lambda/2\right)} +  \frac{\lambda_0 - \Delta \lambda/2}{\left(\lambda_0 - \Delta \lambda/2 -\lambda_0 - \Delta \lambda/2\right)}\right)
\end{equation}

\begin{equation}
\Delta \phi_1 - \Delta \phi_2 = 2\pi \left(\frac{\lambda_0 + \Delta \lambda/2}{\Delta \lambda} +  \frac{\lambda_0 - \Delta \lambda/2}{- \Delta \lambda}\right)
\end{equation}

\begin{equation}
\Delta \phi_1 - \Delta \phi_2 = 2\pi \left(\frac{\lambda_0 + \Delta \lambda/2}{\Delta \lambda} -  \frac{\lambda_0 - \Delta \lambda/2}{\Delta \lambda}\right)
\end{equation}

\begin{equation}
\Delta \phi_1 - \Delta \phi_2 = 2\pi
\end{equation}

This then shows that if the wavelength is kept within the spectral bandwidth, the maximum additional phase shift introduced between two paths will be $2\pi$, validating the assumption made when taking the matrix logarithm in Section \ref{subsec:firstordermodel}.

Intuitively, if path lengths vary in phase by  $\ll 2 \pi$ when the wavelength is varied, the diffracted far-field pattern will remain similar and so the `speckle correlation' will be high.  When the path length differs by $> 2 \pi$, variations in wavelengths produce quite different diffraction patterns (as the phases will be significantly altered) and thus the speckle patterns will become decorrelated.  This is described in more detail in \cite{Mosk2012}.

\section{Bandwidth of step-index MMF}
\label{sec:MMFBWcalc}

To validate the first-order recovery algorithm, we use a dataset consisting of TMs taken at multiple wavelengths (1525--1567nm in steps of 0.08nm) \cite{Carpenter2016a}.  The TMs are taken from a step-index refractive index profile MMF with 420 modes (including polarisation modes).  We first examine the bandwidth over which the first-order model is approximately accurate.  To do this, we consider the TM as a function of wavelength, $\mathbf{A}_{\mathrm{MMF}}(\lambda)$, where $\lambda$ is wavelength in nm.  Using a measured TM at a reference wavelength, $\lambda_{ref}$ (1526nm is used here), we estimate TM at some test wavelength, $\lambda_{test} > \lambda_{ref}$ as:

\begin{equation}
\mathbf{A}_{est}(\lambda_{test}, \lambda_{ref}) = e^{\left(\frac{\lambda_{ref}}{\lambda_{test}} \log \mathbf{A}_{\mathrm{MMF}}(\lambda_{ref})\right)}
\end{equation}

We then compute an error metric to test the predictive accuracy of the first-order model:

\begin{equation}
\label{eq:bwMatrixErrorApp}
\epsilon_{est}(\lambda_{test}, \lambda_{ref}) = \sqrt{\frac{\left \lVert \mathbf{A}_{est}(\lambda_{test}, \lambda_{ref}) - \mathbf{A}_{\mathrm{MMF}}(\lambda_{test}) \right \rVert_F}{\left \lVert \mathbf{A}_{\mathrm{MMF}}(\lambda_{test})\right \rVert_F}}
\end{equation}

\noindent where $\mathbf{A}_{\mathrm{MMF}}(\lambda_{test})$ is the measured MMF TM at wavelength $\lambda_{test}$ and $\left \lVert .. \right \rVert_F$ is the Frobenius norm. Defining $\Delta \lambda = \lambda_{test} - \lambda_{ref}$ we can rewrite Equation \ref{eq:bwMatrixErrorApp} as:

\begin{equation}
\epsilon_{est}(\Delta \lambda) = \sqrt{\frac{\left \lVert \mathbf{A}_{est}(\Delta \lambda, \lambda_{ref}) - \mathbf{A}_{\mathrm{MMF}}(\lambda_{ref} + \Delta \lambda) \right \rVert_2}{\left \lVert \mathbf{A}_{\mathrm{MMF}}(\lambda_{ref} + \Delta \lambda)\right \rVert_2}}
\end{equation}

\begin{figure}[htpb]
	\centering
	\includegraphics[height=0.35\textheight]{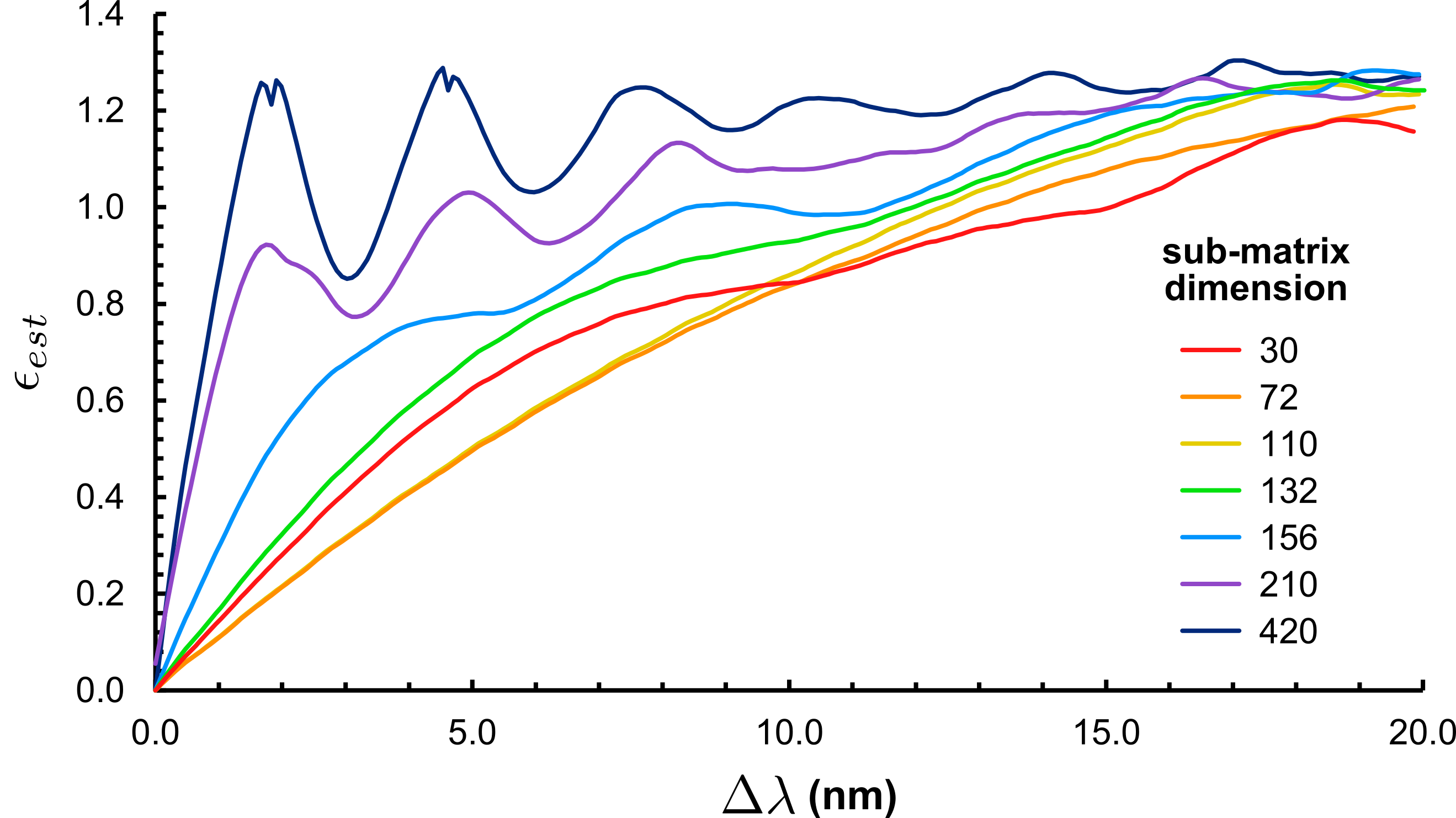}
	\caption{\textbf{Validation of the first-order model of Section \ref{subsec:expSetupAppendix} using measured multi-wavelength TMs from a step-index MMF \cite{Carpenter2016a}.}  The normalised error of the TM predicted by the first-order model compared with the actual measured TM ($\epsilon_{est}$) is plotted as the difference in reference and test wavelengths ($\Delta \lambda$) is increased.  Different sub-matrices of only the lowest-order modes are considered: that is, sub-matrices of size $N \times N$ with $N = 30, 72, 110, 132, 156, 210, 420$.  For lower-order sub-matrices the first-order model is seen to be valid over a larger wavelength range.}
	\label{fig:MMFBWgraph}
\end{figure}

Setting $\lambda_{ref}$ to 1526nm, we can then plot $\epsilon_{est}$ as a function of $\Delta \lambda$ (Figure \ref{fig:MMFBWgraph}).  We find empirically that changing $\lambda_{ref}$ does not significantly change this curve.

The full $420 \times 420$ matrix provides $\epsilon_{est} =  0.5$ for a $\Delta \lambda$ of $\sim$0.5nm.  This is insufficient to to enable the design of a reflector stack like that in Section \ref{subsec:implementingReflectors} using off-the-shelf optical filters because the spacings between filter centre wavelengths would need to be $<0.1$nm.  To improve bandwidth, we restrict analysis to the lowest order modes of the fibre and form an $N \times N$ sub-matrix comprising the first $N$ columns of the first $N$ rows of each $\mathbf{A}_{\mathrm{MMF}}$.  The resulting $\epsilon_{est}$ curves for $N = 30, 72, 110, 132, 156, 210, 420$ are plotted in Figure \ref{fig:MMFBWgraph}.  We find that using the 110 lowest-order modes gives a $\Delta \lambda$ of 5nm for $\epsilon_{est} =  0.5$.  This is ample to allow design of a realistic filter stack in line with the principles of Section \ref{subsec:implementingReflectors} and \ref{subsec:selectingWavelengths} and so is used here.

In reality dispersion will further change the effective optical thicknesses at the different wavelengths.  From the dataset here, we compute that over a 5nm range the error introduced is $<$3\% and so is therefore neglected in this work.


%


\end{document}